\documentclass[twocolumn,tighten]{aastex7}

\newcommand{\beq}[1]{\begin{equation}\label{#1}}
\newcommand{\eeq}{\end{equation}}
\newcommand{\beqn}[1]{\begin{eqnarray}\label{#1}}
\newcommand{\eeqn}{\end{eqnarray}}
\newcommand{\sub}[1]{_\mathrm{#1}}

\newcommand{\tage}{t\sub{\star, age}}
\newcommand{\Op}{\Omega\sub{p}}
\newcommand{\Os}{\Omega\sub{\star}}
\newcommand{\Obar}{\Bar{\Omega}\sub{\star}}
\newcommand{\Ocri}{\Omega\sub{critical}}
\newcommand{\Rc}{R\sub{c}}
\newcommand{\Mc}{M\sub{c}}

\newcommand{\gcons}{\mathcal{G}}

\newcommand{\Rp}{R\sub{p}}

\newcommand{\Mp}{M\sub{p}}

\newcommand{\Mprate}{\dot{M}\sub{p}}

\newcommand{\Mjup}{\mathrm{M}\sub{J}}
\newcommand{\Rjup}{\mathrm{R}\sub{J}}
\newcommand{\Msun}{\mathrm{M}_{\odot}}
\newcommand{\Rsun}{\mathrm{R}_{\odot}}
\newcommand{\Osun}{\Omega_{\odot}}

\newcommand{\Mstar}{M\sub{\star}}

\newcommand{\Rstar}{R\sub{\star}}
\newcommand{\Mstarrate}{\dot{M}\sub{\star}}
\newcommand{\Msunrate}{\dot{M}_{\odot}}

\newcommand{\apos}{a\sub{p}}

\newcommand{\npp}{n\sub{p}}

\newcommand{\Qp}{Q\sub{p}}

\newcommand{\Qs}{Q_\star}

\newcommand{\Ip}{I\sub{p}}
\newcommand{\Is}{I\sub{\star}}
\newcommand{\Lorb}{L\sub{orb}}

\newcommand{\Der}{\mathrm{d}}

\newcommand{\porb}{P\sub{orb}}
\newcommand{\prot}{P\sub{rot, \star}}

\newcommand{\dpdtide}[2]{\dot{P}\sub{TIDE}=(#1 \pm #2)\,\mathrm{ms}\,\mathrm{yr}^{-1}}
\newcommand{\dpdt}[2]{\dot{P}\sub{OBSV}=(#1 \pm #2)\,\mathrm{ms}\,\mathrm{yr}^{-1}}
\newcommand{\ptide}{\dot{P}_\mathrm{TIDE}}
\newcommand{\pobsv}{\dot{P}_\mathrm{OBSV}}

\newcommand{\epp}{\varepsilon\sub{p}}
\newcommand{\eps}{\varepsilon\sub{\star}}
\newcommand{\gyrp}{\alpha\sub{p}}
\newcommand{\gyrs}{\alpha\sub{\star}}
\newcommand{\omp}{\Dot{\Omega}\sub{wind, p}}
\newcommand{\oms}{\Dot{\Omega}\sub{wind, \star}}

\newcommand{\alphap}{\alpha\sub{p}}

\newcommand{\betap}{\beta\sub{p}}

\newcommand{\gammap}{\gamma\sub{p}}

\newcommand{\alphas}{\alpha\sub{\star}}
\newcommand{\betas}{\beta\sub{\star}}

\newcommand{\gammas}{\gamma\sub{\star}}

\newcommand{\msyr}{\mathrm{ms}\,\mathrm{yr}^{-1}}

\definecolor{mycolor}{rgb}{0.418, 0.188, 0.478}

\usepackage{amsmath}
\usepackage{newtxtext,newtxmath}
\usepackage[T1]{fontenc}
\usepackage{balance}
\usepackage{braket}
\usepackage{tablefootnote}
\usepackage{subfigure}
\usepackage{booktabs}
\usepackage{hyperref}

\makeatletter
\def\hlinewd#1{%
\noalign{\ifnum0=`}\fi\hrule \@height #1 \futurelet
\reserved@a\@xhline}
\makeatother

\shortauthors{Jaime A. Alvarado-Montes et al.}

\begin{document}

\title{Orbital Decay of the Ultra-Hot Jupiter TOI-2109b: Tidal Constraints and Transit-Timing Analysis}

\correspondingauthor{Jaime A. Alvarado-Montes}
\email[show]{jaime.alvaradomontes@mq.edu.au}

\author[0000-0003-0353-9741]{Jaime A. Alvarado-Montes}
\altaffiliation{Macquarie University Research Fellow (MQRF)}
\affiliation{Australian Astronomical Optics, Macquarie University, Balaclava Road, Sydney, NSW 2109, Australia.}
\affiliation{Astrophysics and Space Technologies Research Centre, Macquarie University, Balaclava Road, Sydney, NSW 2109, Australia.}
\email{jaime-andres.alvarado-montes@hdr.mq.edu.au}

\author[0000-0002-8065-4199]{Mario Sucerquia}
\affiliation{Univ. Grenoble Alpes, CNRS, IPAG, 38000 Grenoble, France.}
\email{mario.sucerquia@univ-grenoble-alpes.fr}


\author[0000-0002-6140-3116]{Jorge I. Zuluaga}
\affiliation{SEAP/FACom, Instituto de F\'isica - FCEN, Universidad de Antioquia, Calle 70 No. 52-21, Medell\'in, Colombia.}
\email{jorge.zuluaga@udea.edu.co}

\author[0000-0002-4046-987X]{Christian Schwab}
\affiliation{School of Mathematical and Physical Sciences, Macquarie University, Balaclava Road, Sydney, NSW 2109, Australia.}
\email{christian.schwab@mq.edu.au}

\begin{abstract}
TOI-2109b is the ultra-hot Jupiter with the shortest orbital period ($\sim$16\,hr) yet discovered. At this close distance, strong tidal interactions can produce a significant exchange of angular momentum with the star. Since the orbital period of this planet is shorter than the stellar rotation period, TOI-2109b may be an optimal candidate for studying orbital decay. This process depends on how efficiently the star and the planet dissipate energy, due mainly to interior mechanisms that are poorly constrained in exoplanet systems. In this work, we study for the first time the tidal evolution of TOI-2109b under a formalism of inertial waves (IWs) in convective envelopes and internal gravity waves (IGWs) in stellar radiative regions. We find that uncertainties in the age of TOI-2109 ($\tage$) significantly affect the rate of orbital evolution, as IWs and IGWs interact differently depending on $\tage$. For an `old' host star, we find that TOI-2109b would undergo fast orbital decay. Conversely, if TOI-2109b orbits a `young' host star, a rather slow decay rate for $Q\sub{\star}'>2.3\times10^7$ would suggest a constant-period orbit. Our calculated mid-transit times and transit-timing variations (TTVs) support a `young' host star with $Q\sub{\star}'>3.7\times10^7$, suggesting a decay rate $\dot{P}\sim4\,\msyr$ that could lead to mid-transit-time shifts $\lesssim10$\,s over three years. Orbital decay and other TTV-inducing effects will be confirmed or ruled out with future higher-quality timing data. The results presented here aim at constraining the current modeling of tides and TTVs for TOI-2109b, helping us further understand light-curve changes associated to the long-term evolution of ultra-short-period planets.
\end{abstract}




\section{Introduction}
\label{sec:intro}

\begin{figure}
    \centering
    \includegraphics[scale=0.57]{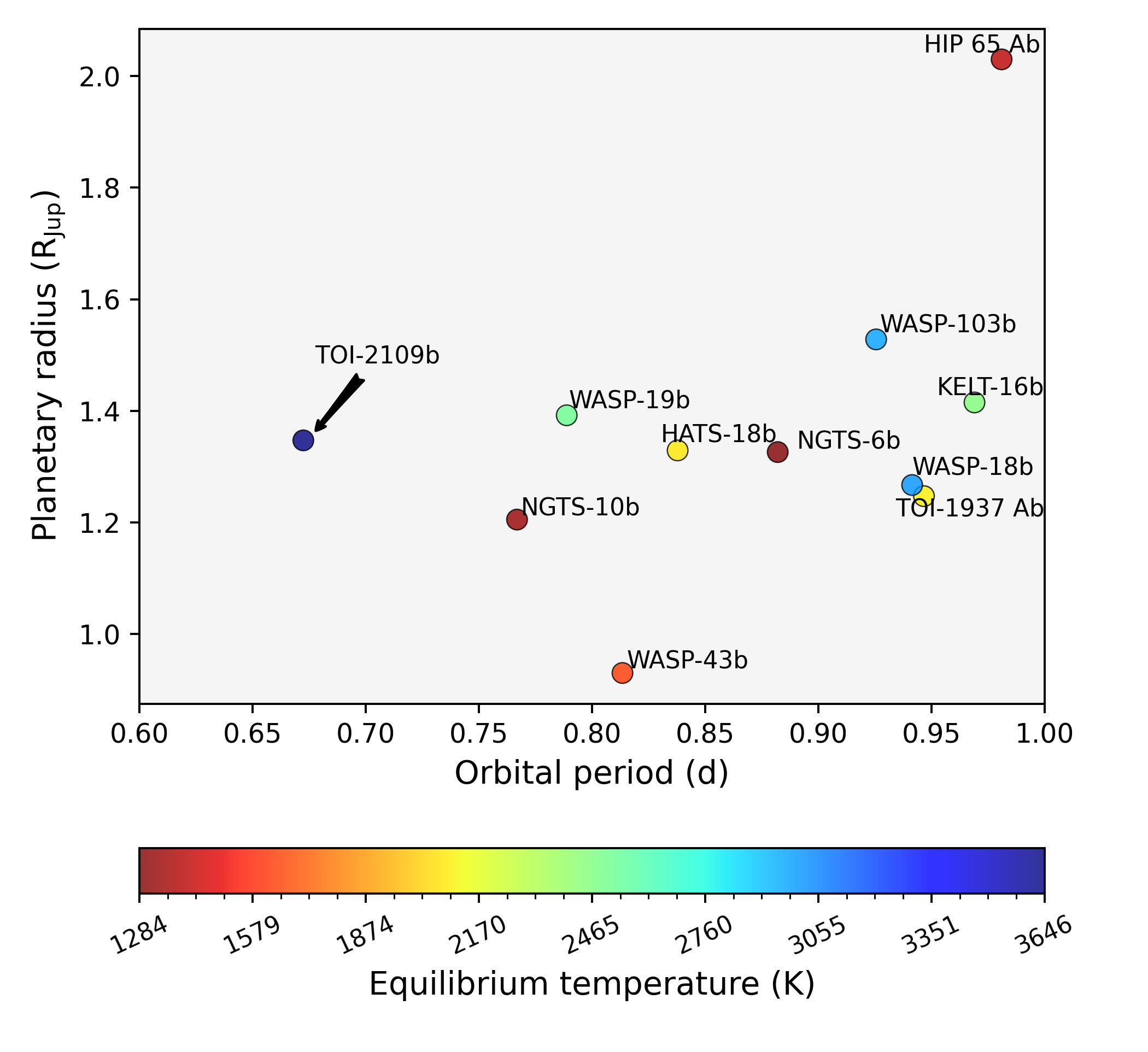}
    \caption{Distribution of USP Jupiter-like planets \citep[$\porb<1$ d;][]{Winn2018}. Colors represent equilibrium temperatures. Data extracted from the \citet{NASAExoArch}.}
    \label{fig:distri}
\end{figure}

A low occurrence rate of $\sim0.5\%$ suggests that hot Jupiters (HJs) orbiting Sun-like stars are extremely rare \citep[see, for example,][]{Howard2012,Wright2012}. Since the discovery of the first of these systems \citep{Mayor1995}, HJs have significantly modified and expanded our knowledge of planet formation, evolution, and dynamics \citep{Dawson2018}. Particularly, a handful of these massive giants are located in the most extreme orbits: the so-called ultra-short-period Jupiters (USP-Js; Figure~\ref{fig:distri}) with orbital periods $\porb<1$ d \citep{Winn2018} or possibly $\porb<2$ d \citep{Goyal2025}. In such locations, the intense gravitational/tidal interactions of these planets can induce a bulge in the host star, producing torques that can trigger a transfer of angular momentum with the planetary orbit. For spin-orbit alignment, a significant transfer of angular momentum can lead to tidal orbital decay \citep[e.g.,][]{Ma2021, Alvarado-Montes2022}. In USP-J systems, orbital decay is an impending yet elusive outcome \citep{Petrucci2020,Adams2024} that is ultimately determined by the planetary mass \citep{Hamer2020}, and it offers a unique opportunity to probe tidal evolution models.

Tidal orbital decay only occurs if the stellar rotational rate is slower than the planetary orbital rate \footnote{If, on the contrary, the star rotates faster than the planet moves around the star, the planet may potentially migrate outward and spin down the star \citep[e.g.,][]{Cohen2010, Kotorashvili2024}} \citep{Penev2018}. In this scenario, the planet moves forward and the tidal bulge is dragged behind the sub-planetary point on the stellar surface.\footnote{Except when stellar spin and planet orbit are synchronized \citep{Hut1981}.} Consequently, the planetary orbit may shrink \citep{Rosario2022} and the star may spin up \citep{Lanza2010,Brown2011,Penev2016}. Orbital decay in USP-Js can produce planetary mass loss \citep{Valsecchi2015,Jackson2016}, tidal deformation of the planet's equilibrium shape \citep{Wolf2009,Akinsanmi2019,Wahl2021,Barros2022, Akinsanmi2024}, anomalies in measured planetary radii \citep{Leconte2010,Millholland2019,MolLous2020,Hou2022,Rozner2022}, and disruption due to strong stellar tides \citep{Levrard2009,Delrez2016,Oberst2017}. All these phenomena confirm the need of improving our understanding of tidal interactions and orbital dynamics.

As the planet dissipates orbital kinetic energy due to friction inside the star, tidal orbital decay is produced. This process can then help us understand how stars dissipate internal energy \citep[e.g.,][]{Fellay2023}. The efficiency of that dissipation is quantified via the tidal quality factor $Q_\star' = 3Q_\star/(2k_{2, \star})$ \citep{Goldreich1966}, a highly unknown parameter that changes according to various stellar properties \citep[see, e.g,][]{Barker2020}. For stars, theoretical values range from $10^5$ to $10^9$, where smaller values indicate higher efficiencies. $Q\sub{\star}'$ strongly depends on the tidal forcing period and amplitude, and for $0.5<\porb<2$ d it lies between $10^5$ and $10^7$ \citep{Penev2018}. Thus, tidal orbital decay in USP-Js can be used to constrain $Q\sub{\star}'$ \citep[e.g.,][]{Yee2020,Vissapragada2022,Tokuno2024,Harre2024}, so stronger evidence is needed to probe orbital decay in these systems \citep{Patra2020}.

The above can be accomplished using transit-timing measurements, a.k.a. transit-timing variations (TTVs). To measure TTVs in a transiting exoplanet, a dedicated and precise photometric follow-up is needed \citep[e.g.,][]{Yeh2024}. WASP-12b is a good example of this, whose changes in the orbital period were initially detected by \citet{Maciejewsky2016} and confirmed by \citet{Patra2017,Bailey2019}. Several TTV studies have found that such changes are due to WASP-12b's orbital decay \citep{Yee2020,Wong2022,Hagey2022,Leonardi2024}. Similarly, using {\it Kepler} and {\it TESS} data, \citet{Vissapragada2022} presented evidence of orbital decay in the Kepler-1658 system that might be tidally-induced \citep{Barker2024}. \citet{Bouma2019} suggested that WASP-4b may also be experiencing orbital decay, albeit more compelling evidence is required to support this idea as other studies attribute the observed changes to the R\o mer effect\footnote{Line-of-sight acceleration due to wide stellar companions \citep{Harre2023}}. More exoplanet systems have also been analysed for orbital decay \citep{Biswas2024} and the possible acceleration of said decay \citep{AlvaradoEF2024}.

USP-Js carry fruitful information about stellar internal structures and their energy dissipation mechanisms \citep{Essick2016,Barker2020}. These include the dissipation of equilibrium tides due to turbulent convection \citep{Zahn1989} or instabilities in convective \citep[e.g.,][]{Barker2016} and radiative regions \citep[e.g.,][]{Vidal2018}, the excitation of inertial waves (IWs) restored by Coriolis forces in stellar convective envelopes \citep{Ogilvie2007,Rieutord2010,Ogilvie2013, Mathis2015,Bolmont2016,Gallet2017,Benbakoura2019}, and the excitation of internal gravity waves (IGWs) via tidal forcing in stellar radiative regions \citep{Barker2010,Barker2011,Weinberg2012,Essick2016,Lazovik2021}. The study of these and other mechanisms\footnote{For example, rare processes such as the conversion of IGWs into magnetic waves may play a crucial role in how stars with strong magnetic fields dissipate energy \citep{Lin2018,Duguid2024}.}, in turn, can help us further understand how variations in the stellar physical properties affect the architecture of solar systems.

TOI-2109b is a USP-J characterized by \citet{Wong2021} with an orbital period of $\sim16$ hours, initially identified by The Transiting Exoplanet Survey Satellite \citep[{\it TESS};][]{Ricker2015}. TOI-2109b is the shortest-period and second hottest HJ with $T\sub{eq}=3646\pm88$ K (see Figure~\ref{fig:distri}), just behind KELT-9b \citep[$T\sub{eq}=3921^{+182}_{-174}$;][]{Gaudi2017,Jones2022}. The extreme proximity of TOI-2109b from its host star---only about 2.3 stellar radii---is why its predicted orbital decay rate could be the highest of all USP-Js \citep{Wong2021}. However, TOI-2109 is an F-star where shallower convection zones are expected, which combined with its fast rotational rate ($v\mathrm{sin}\,i=81.2\pm1.6$$\,$km$\,$s$^{-1}$) could decrease the efficiency of tidal interactions \citep[e.g.,][]{Harre2023}. In this work, we study the possibility of tidal orbital decay in TOI-2109b by simulating and constraining the interactions that drive the angular momentum exchange via the dissipation of IWs in the convective envelope of the star and the planet, as well as IGWs in the stellar radiative regions. As the population of USP-Js grows, the study of systems such as TOI-2109 can teach us not only about the fate of extreme gas giants, but also about the stellar and planetary properties that may determine the orbital demise of short-period planets.

The layout of this paper is as follows. In Section~\ref{sec:eqs}, we introduce the tidal evolution equations that we use to study the orbital evolution of TOI-2109b along with the caveats of this system. In Section~\ref{sec:dissmech}, we present the stellar and planetary tidal dissipation mechanisms coupled to the evolution model, and in Section~\ref{sec:orbdecay} we present the results of our dynamical simulations of TOI-2109b. In Section~\ref{sec:tt_analysis}, we perform a TTV analysis on available photometric data, constrain the decay rate of TOI-2109b, and compare it with our theoretical results. Finally, in Section~\ref{sec:disc} we discuss the evolution of TOI-2109b in terms of observational predictions and the scope of our study.

\begin{figure*}
    \centering
    \hspace{-0.4cm}
    \includegraphics[width=\textwidth]{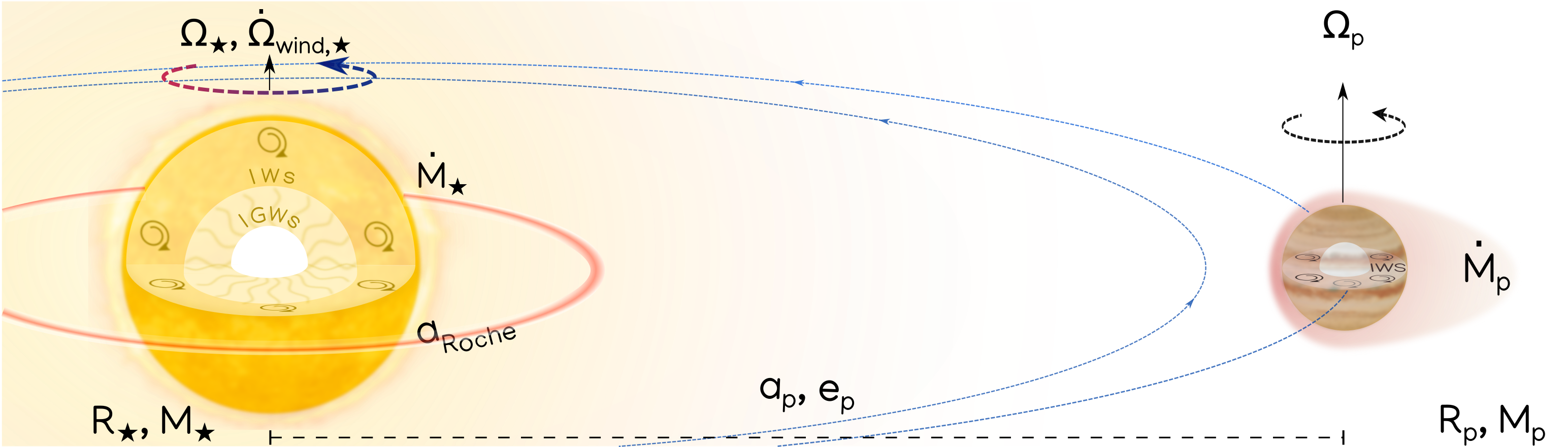}%
    \caption{Schematic depiction of the TOI-2109b system, illustrating the primary physical and orbital parameters that describe the evolution of the system in response to tidal interactions between the planet and its host star.}
    \label{fig:scheme}
\end{figure*}

\section{Tidal Evolution Equations}
\label{sec:eqs}
The mutual stellar and planetary tidal bulges produce torques that affect the planet's orbit and change the rotation of the star. Here, we assume that the stellar rotational axis is completely perpendicular to the orbital plane, an arrangement known as spin-orbit alignment\footnote{For cases where the spin-orbit angle (i.e. obliquity) evolves, frameworks such as that in \citet{Barker2009} must be used.} (see Figure~\ref{fig:scheme} and Section~\ref{sec:orbdecay} for a further explanation on this assumption).

For a star and a planet with mass $\Mstar$ and $\Mp$, respectively, where $\Mp\ll\Mstar$, the evolution of the stellar (planetary) spin, $\Os$ ($\Op$); and the planet mean motion ($\npp$) and eccentricity ($e$) will obey four coupled differential equations. These equations assume the constant tidal lag-time\footnote{This is the delay, $\tau$, of the stellar tidal bulge when compared to the position vector of the planet.} model of \citet{Alexander1973,Hut1981} derived from the equilibrium tide theory presented in \citet{Darwin1879}. The reader should refer to Section~\ref{sec:disc} for a discussion on this chosen model. Such equations are as follows and apply for any eccentricity:

\beq{eq:dnpdt}
\frac{\dot{n}\sub{p}}{\npp} = -3\left[\frac{\dot{e}}{e}\frac{e^2}{1-e^2}-\frac{\Is\dot{\Omega}_\star}{\Lorb}-\frac{\Ip\dot{\Omega}\sub{p}}{\Lorb}+\frac{\Is\oms}{\Lorb}\right],
\eeq
\beq{eq:dedt}
\begin{split}
\frac{\dot{e}}{e}=\frac{27\npp}{\apos^5}&\bigg\{\frac{3}{2\Qp'}\frac{\Mstar}{\Mp}\Rp^5\left[\frac{11}{18}\frac{\Op}{\npp}e_2(e)-e_1(e)\right]\\&+\frac{3}{2\Qs'}\frac{\Mp}{\Mstar}\Rstar^5\left[\frac{11}{18}\frac{\Os}{\npp}e_2(e)-e_1(e)\right]\bigg\},
\end{split}
\eeq
\beq{eq:dopdt}
\frac{\Der\Op}{\Der t}= \frac{9\npp^4\Rp^3}{2\epp\gyrp \gcons\Mp\Qp'}\left[e_3(e)-e_4(e)\left(\frac{\Op}{\npp}\right)\right] +\,\omp,
\eeq
\beq{eq:dosdt}
\frac{\Der\Os}{\Der t}= \frac{9\npp^4\Mp^2}{2\eps\gyrs \gcons\Qs'}\left(\frac{\Rstar}{\Mstar}\right)^3\left[e_3(e)-e_4(e)\left(\frac{\Os}{\npp}\right)\right] +\,\oms,
\eeq
where $e_1$, $e_2$, $e_3$, and $e_4$ are functions of $e$:
\begin{align}
\label{eq:ecc1}
e_1(e) &= \left(1 + \frac{15}{4}e^2 + \frac{15}{8}e^4 + \frac{5}{64}e^6\right) \bigg/ (1-e^2)^{13/2},\\
\label{eq:ecc2}
e_2(e) &= \left(1 + \frac{3}{2}e^2 + \frac{1}{8}e^4\right) \bigg/ (1-e^2)^{5},\\
\label{eq:ecc3}
e_3(e) &= \left(1 + \frac{15}{2}e^2 + \frac{45}{8}e^4 + \frac{5}{16}e^6\right) \bigg/ (1-e^2)^6,\\
\label{eq:ecc4}
e_4(e) &= \left(1 + 3e^2 + \frac{3}{8}e^4\right) \bigg/ (1-e^2)^{9/2}.
\end{align}

If the star (planet) has a radius $\Rstar$ ($\Rp$), the stellar (planetary) moment of inertia will be $\Is = \eps\gyrs\Mstar\Rstar^2$ ($\Ip = \epp\gyrp\Mp\Rp^2$). The planet's orbital angular momentum, $\Lorb$, is given by
\beq{eq:lorb}
\Lorb = \Mp\Mstar\sqrt{\frac{\gcons\apos(1-e^2)}{\Mstar + \Mp}},
\eeq
where $\gcons$ is the gravitational constant and $\apos = (G\Mstar/\npp^2)^{1/3}$ is the planet semimajor axis. The coefficients $\alpha\sub{\star, p}$ are determined by the stellar and planetary interior structures (Equation~\ref{eq:k2Qparameters}). The mass fraction participating in the exchange of angular momentum is set as $\varepsilon\sub{\star, p}\simeq1$, as both star and planet have extensive convective regions \citep{Gu2003}.

The evolution of the system is affected by stellar wind torques ($\oms$ in Equations~\ref{eq:dnpdt} and~\ref{eq:dosdt}). For the case considered here, we neglect planetary winds ($\omp$ in Equation~\ref{eq:dopdt}). The loss of angular momentum due to magnetic braking and the loss of stellar mass will be estimated following the formalism by \citet{Weber1967} and \citet{Johnstone2015}:
\begin{align}
\label{eq:magbrak}
\oms&=-\kappa \Os \mathrm{Min}(\Os, \Obar)^2,~\mathrm{and}\\
\label{eq:stellarmassrate}
\Mstarrate&=\left(\frac{\Mstar}{\Msun}\right)^{a}
\left(\frac{\Os}{\Osun}\right)^{b} \left(\frac{\Rstar}{\Rsun}\right)^{2}\Msunrate,
\end{align}
respectively. In Equation~(\ref{eq:magbrak}), $\Obar$ stands for the `saturation' rate: an upper limit where magnetic braking is no longer dependent on the stellar spin. Also, $\kappa$ determines the physical scaling of magnetic braking as $\kappa = \Os^{-2}/2\tage$, with $\tage$ the stellar age \citep[e.g.,][]{Finley2017}. In Equation~(\ref{eq:stellarmassrate}), $a=-3.36$, $b=1.33$, $\Osun=2.67\times10^{-6}$~rad\,s$^{-1}$, and $\Msunrate=1.4\times10^{-14}\,\Msun$~yr$^{-1}$.

The planetary mass will decrease for photo-evaporation and stellar wind drag, and it will be calculated using a power-law as \citep{Zendejas2010,Sanz-Forcada2011}:
\beq{eq:Mloss-rate}
\Mprate = -\frac{\pi \Rp^{3} F_\mathrm{XUV}}{\gcons K\Mp}- \left(\frac{\Rp}{\apos}\right)^2 \frac{\Mstarrate \alpha}{2},
\eeq
where $F_\mathrm{XUV}$ is the X-ray and extreme UV stellar flux, $\alpha=0.3$ is an entrainment efficiency factor, and $K$ accounts for the planet radius losses at the Roche lobe.

In Equations~(\ref{eq:dnpdt})--(\ref{eq:dosdt}), $Q\sub{\star,\,p}'$ represents the tidal energy dissipation of the star ($\star$) and the planet ($\mathrm{p}$). The functional form of $Q'$ is complicated owing to its dependence on the tidal frequency $\omega$ \citep[see, e.g.,][]{Ogilvie2007}, which is a result of the intricate internal mechanisms whereby interacting rotating bodies dissipate energy \citep{Zahn2008}. Furthermore, the arbitrary eccentricities and obliquities of extrasolar systems can make the coupling of $Q'$ into the evolution of orbital parameters even more difficult. For these reasons, to study tidal dissipation efficiency we follow a typical assumption where $Q'$ is inversely proportional to $\omega$, that is, $Q'=1/\omega\tau$ \citep[e.g.,][]{Barker2009}. To include this into the tidal evolution equations presented before, and in the same spirit of \citet{Alvarado-Montes2022}, we adopt a constant-$Q'$ model where $\omega\simeq2\npp$ for the dominant harmonic of the tide, so that $Q'=1/(2\npp\tau)$. Such an assumption entails: direct proportionality of $\tau$ with the orbital period of the planet, independence of $\omega$, and homogeneity for all the components of the tide.


\section{Energy Dissipation Mechanisms}
\label{sec:dissmech}

The energy dissipation of the star and the planet (and its coupling into the planet's orbital evolution) is encoded in the time dependence of their $Q'$, for which we adopted a piece-wise homogeneous two-layer model for each body. Detailed expressions for this quantity can be found in \citet{Ogilvie2013,Guenel2014,Mathis2015}, and have been well tested in different systems \citep[e.g.,][]{Alvarado2019,Barker2020,Alvarado-Montes2022}. We model the internal energy dissipation of the host star TOI-2109 via dynamical waves: inertial waves (IWs) in the convective envelope and internal gravity waves (IGWs) in the radiative regions. The energy dissipated by the planet was also modeled through the planetary tidal quality factor (i.e. $\Qp'$), including inertial waves in the convective envelope of the planet as well as the viscoelasticity of its solid core.

\subsection{Inertial waves (IWs) in the stellar (planetary) convective envelope of TOI-2109 (b)}
\label{sec:k2q_IW}

The inertial terms in the equation of motion contain the dynamical tide that is forced by the equilibrium tide in convective zones. However, when rotation is present the effect of tides is strongly frequency-dependent, making it a problem harder to solve \citep[e.g.,][]{Rieutord2010}.

Tidal interactions between two or more bodies are complex and can affect the equilibrium shape of interacting bodies. Such tidal deformations in a star and planet can be assumed proportional to the their force if the amplitude of the perturbations is small \citep{Love1927}, and their response to tidal distortion can be quantified via intricate coefficients that, again, strongly depend on the tidal frequency, $\omega$ \citep[e.g., ][]{Efroimsky2012}. Those coefficients are the so-called Love numbers, $k_l^m(\omega)$, for which the imaginary part of the $l=m=2$ component of the tide, $k_2^2(\omega)$, provides information about the tidal dissipation via IWs\footnote{Assuming that the obliquities of the star and the planet are negligible.},
\beq{eq:Imk2}
\frac{2}{3Q'} =\int_{-\infty}^{+\infty}{\rm Im}[k_2(\omega)]\frac{\Der\omega}{\omega}=\int_{-\infty}^{+\infty}\frac{|k_2^2(\omega)|}{Q_2^2(\omega)}\frac{\Der\omega}{\omega}.
\eeq

Similar to the formalism of \citet{Ogilvie2013} that uses an impulsive forcing to calculate IWs, we use here a frequency-averaged formalism to compute the tidal excitation of these waves in convection zones and quantify their typical dissipation. Thus, the previous equation is averaged in $\omega\,\epsilon\,[-2\Omega, 2\Omega]$. Prior results have shown a more detailed inclusion of the realistic interior structure of a body \citep[see e.g., ][]{Gallet2017, Benbakoura2019}. However, using a frequency-averaged dissipation is a coarse estimate that allows us to approach such an intricate problem by providing us a typical quantification of the dissipation produced by IWs. The actual quantification of IW-dissipation\footnote{This depends, for example, on the combination of non-linear effects with magnetic fields and/or turbulence in convection zones. This is not addressed here as studying the specific dissipation mechanism of IWs is out of the scope of this work.} for a specific tidal frequency in an individual system is subject to large uncertainties (from 2 to 3 orders of magnitude), which is why we use a formalism that is robust and has been widely adopted in previous research.

As shown by some authors \citep[e.g.,][]{Ogilvie2013,Barker2020}, tidal dissipation due to IWs is significantly more efficient for rapid rotators and will only apply if $|\omega|\leq2\Os$ (i.e. while tidal forcing excites IWs), which for circular orbits or small eccentricities will be true only if $\porb\geq\frac{\prot}{2}$. Seeking a robust (but straightforward) implementation of tidal dissipation via IWs into the evolution of the TOI-2109 system and using the aforementioned frequency-averaged formalism, we employ a seemingly simple numerical implementation. This is performed with a simplified piece-wise homogeneous two-layer model for both the star (subscript $\star$) and the planet (subscript p). For the star, a fluid-fluid boundary is used between the two layers \citep{Mathis2015}, whereas for the planet it is a solid-fluid boundary \citep{Guenel2014},
\beq{eq:k2QFormulas}
\begin{split}
    \frac{1}{\braket{ Q\sub{\star,\,IW}'}} =& \frac{200\pi}{189}\epsilon_\mathrm{\Omega,\,\star}^{2}\frac{\alphas^{5}}{1-\alphas^{5}}(1-\gammas^2)(1-\alphas^2)\times\\&\left(1 + 2\alphas + 3\alphas^2 + \frac{3}{2}\alphas^3\right)^2\left[1+\left(\frac{1-\gammas}{\gammas}\right)\alphas^3\right]\times\\&\bigg[1 + \frac{3}{2}\gammas+\frac{5}{2\gammas}\left(1 + \frac{1}{2}\gammas-\frac{3}{2}\gammas^2\right)\alphas^3-\\&\frac{9}{4}(1-\gammas)\alphas^5\bigg]^{^{-2}}
\end{split}
\eeq
\beq{eq:k2QFormulap}
\begin{split}
    \frac{1}{\braket{Q\sub{p}'}} =& \frac{200\pi}{189}\epsilon_\mathrm{\Omega,\,p}^{2}\frac{\alphap^{5}}{1-\alphap^{5}}\left[1+
	\frac{1-\gammap}{\gammap}\alphap^{3}\right]\times\\&\left[1+
		\frac{5}{2}\frac{1-\gammap}{\gammap}\alphap^{3}\right]^{^{-2}} +\,\frac{\pi\mathcal{R}(3 + \mathcal{A})^{2}\mathcal{BC}}{\mathcal{D}(6\mathcal{D} + 4\mathcal{ABCR})},
\end{split}
\eeq
where $\epsilon_\Omega^{2}\equiv (\Omega/\Ocri)^{2}$ and $\Ocri\equiv(\gcons M/R^{3})^{1/2}$. Hereafter, $\braket{ Q\sub{\star,\,IW}'}\equiv Q\sub{\star,\,IW}'$ and $\braket{ Q\sub{p}'}\equiv Q\sub{p}'$.

In the case of the planet, the first term in Equation~(\ref{eq:k2QFormulap}) stands for the dissipation via IWs in the convective envelope, whereas the viscoelastic dissipation of the inner solid core is represented by the second term. The rigidity of the planet's core, $\mathcal{R}$, is given in Pascals [Pa]. The auxiliary functions $\mathcal{A}$, $\mathcal{B}$, $\mathcal{C}$, and $\mathcal{D}$ are given by
\beq{eq:corepars}
\begin{split}
&\mathcal{A} = 1 + \frac{5}{2}\gamma^{-1}\alpha^3(1-\gamma), \quad\mathcal{B} = \alpha^{-5}(1 - \gamma)^{-2},\\&
\mathcal{C} = \frac{38\pi}{3}\frac{(\alpha\Rp)^4}{\gcons(\beta\Mp)^2},\hspace{0.12cm}\mathcal{D} = \frac{2}{3}\mathcal{A}\mathcal{B}(1-\gamma)\left(1+\frac{3}{2}\gamma\right) - \frac{3}{2}.
\end{split}
\hspace{0.8cm}
\eeq
\indent Dissipation resulting from IWs in convection zones is strongly dependent on the interior structure of the star and the planet (see, e.g., $Q\sub{\star,\,IW}'$ in Figure~\ref{fig:contourplot}). Such a dependence is evident through Equations~(\ref{eq:k2QFormulas})--(\ref{eq:corepars}) where $\alpha$ and $\beta$ are defined as the size and mass aspect ratios of the radiative/convective (star) or solid/convective (planet) interfaces,
\beq{eq:k2Qparameters}
\alpha \equiv \frac{R_\mathrm{c}}{R},\quad
\beta \equiv \frac{M_\mathrm{c}}{M},\quad
\eeq
with $\Mc$ and $\Rc$ the core's mass and radius, respectively. The ratio of the mean densities between the two regions for the star and/or planet is defined as,
\beq{eq:k2Qgamma}
\gamma \equiv \frac{\alpha^{3}(1-\beta)}{\beta(1-\alpha^{3})}.
\hspace{1.25cm}
\eeq

Furthermore, \citet{Barker2020} performed an in-depth study of the dissipation efficiency via IWs for different stellar spectral types, leading to various findings about how these waves are excited in convection zones as well as for which scenarios their dissipation is more dominant. Their modeling, though mathematically speaking does not represent a much higher level of complexity, is more computationally 
demanding. While it may offer a better representation of stellar physical processes, the simplicity, availability, and widely-adopted use of the two-layer model make it more accessible to a wider range of calculations. However, a detailed comparison of the results achieved for IW-dissipation using a realistic model and a two-layer model showed that the latter underpredicts the dissipation of IWs in F-type stars.

For TOI-2109 where $\Mstar\approx1.4\,\Msun$ and $\tage$ varies from 1 to 2.5 Gyr, such a comparative analysis reveals that $\Qs'$ follows the same qualitative evolution as times goes on, but IW-dissipation is underpredicted in 1--2 order of magnitude in the two-layer model \citep[see fig.~7 in][]{Barker2020}. However, it is worth noting that we use the two-layer model not only to quantify IW-dissipation, but also to include the time dependence of $\Qs'$ through $\Os$, which is directly coupled to the evolution equations. Therefore, we account for the quantitative differences of $\Qs'$ in the two-layer model by using \citet{Barker2020} to scale the value of $\Qs'$ to the expected level of dissipation in F-type stars such as TOI-2109. By keeping this semi-analytical track of how $\Qs'$ changes over time, this approach makes our results easy to be reproduced using the same (or similar) computational methods.

\subsection{Internal gravity waves (IGWs) in the stellar radiative regions of TOI-2109}
\label{sec:k2q_IGW}
The excitation of IWs by tidal forcing in the convective envelope of TOI-2109 (subsection~\ref{sec:k2q_IW}) is not the only dissipation mechanism acting in its interior structure. In fact, as IWs in convective envelopes are excited only if $\porb\geq\frac{\prot}{2}$, once this condition does not hold true, for instance, when the planet reaches an inner orbit where $\porb\approx0.5$ d, the star-planet system will continue to dissipate energy only---at least within this model---through the so-called Internal Gravity Waves (IGWs) produced in the stellar convective/radiative interface. Tidal dissipation via IGWs in TOI-2109 will be significant if IGWs are fully damped, which can happen when IGWs travel and break before reaching another radiative region. In between resonances, travelling waves are the most direct and straightforward way of studying the dissipation of energy via IGWs, thus providing their most efficient tidal dissipation. Moreover,  even if resonance exists within tidal forcing, for large enough tidal amplitudes waves can still break and their full damping might still hold true.

Using the formalism presented in \citet{Barker2020}, we study tidal dissipation via IGWs in two scenarios: 1) while $\porb\geq\frac{\prot}{2}$ so that IWs and IGWs act together, and 2) when the planet reaches inner positions where IWs are no longer excited (i.e., $\porb\ngeq\frac{\prot}{2}$), leaving IGWs as the only active dissipation mechanism for shorter orbital periods. Tidal dissipation due to IGWs in radiative regions strongly depends on the orbital/tidal period. Particularly, after complex numerical calculations coupled to stellar evolution models using \textsc{MESA}, \citet{Barker2020} found $Q\sub{\star, \,IGW}'\propto(P_\mathrm{tide}/0.5\mathrm{d})^{\frac{8}{3}}$ for main-sequence stars, while $Q\sub{\star,\,IGW}'\approx10^{8.5}$ for F-stars of $\Mstar\approx1.45\,\Msun$. We must, however, mention that the difference between these two cases strongly depends on the parameters of the host star, entailing some caveats that need to be considered within the model adopted for the dissipation of IGWs. Also, the treatment used here neglects the effects of rotation on IGWs. Still, it is important to note that the inclusion of such effects in the tidal frequency $\omega$ can lead to an enhanced efficiency of tidal dissipation via IGWs. Please refer to \citet{Ivanov2013} for a complete description.

As mentioned before, and further explained in \citet{Barker2020}, tidally excited IGWs will be very efficient if they are in the traveling wave regime due to nonlinear wave breaking in the stellar core. This means that IGWs need to be fully damped, which can happen if radiative diffusion takes place, if large resonant tidal forcing is present \citep{Barker2010,Barker2011}, and/or if IGWs have large tidal amplitudes so that stratification inside the star can be overturned \citep[see][]{Barker2010}. Previously, \citet{Barker2020} found that only the right combination of planetary mass, stellar mass, and stellar age would lead to wave breaking. This condition is enclosed into a mass threshold for the the planet when exceeding the so-called critical mass, $M_\mathrm{crit}$, which in turn depends on the interior structure of the star\footnote{$M_\mathrm{crit}$ varies with $\tage$ and $\Mstar$. See fig. 9 in \citet{Barker2020}.}. The efficiency of IGWs strongly depends on $M_\mathrm{crit}$ and $M_\mathrm{crit}$ significantly varies with $\tage$, which for TOI-2109 is highly uncertain (see Table~\ref{tab:usp}). As a result, $\tage$ is the overarching stellar parameter that determines the possible orbital decay rates of TOI-2109b, splitting the results into two different scenarios, as described in Section~\ref{sec:orbdecay}.

\section{Orbital decay}
\label{sec:orbdecay}

\begin{figure}
    \centering
    \hspace{-0.4cm}
    \includegraphics[width=\columnwidth]{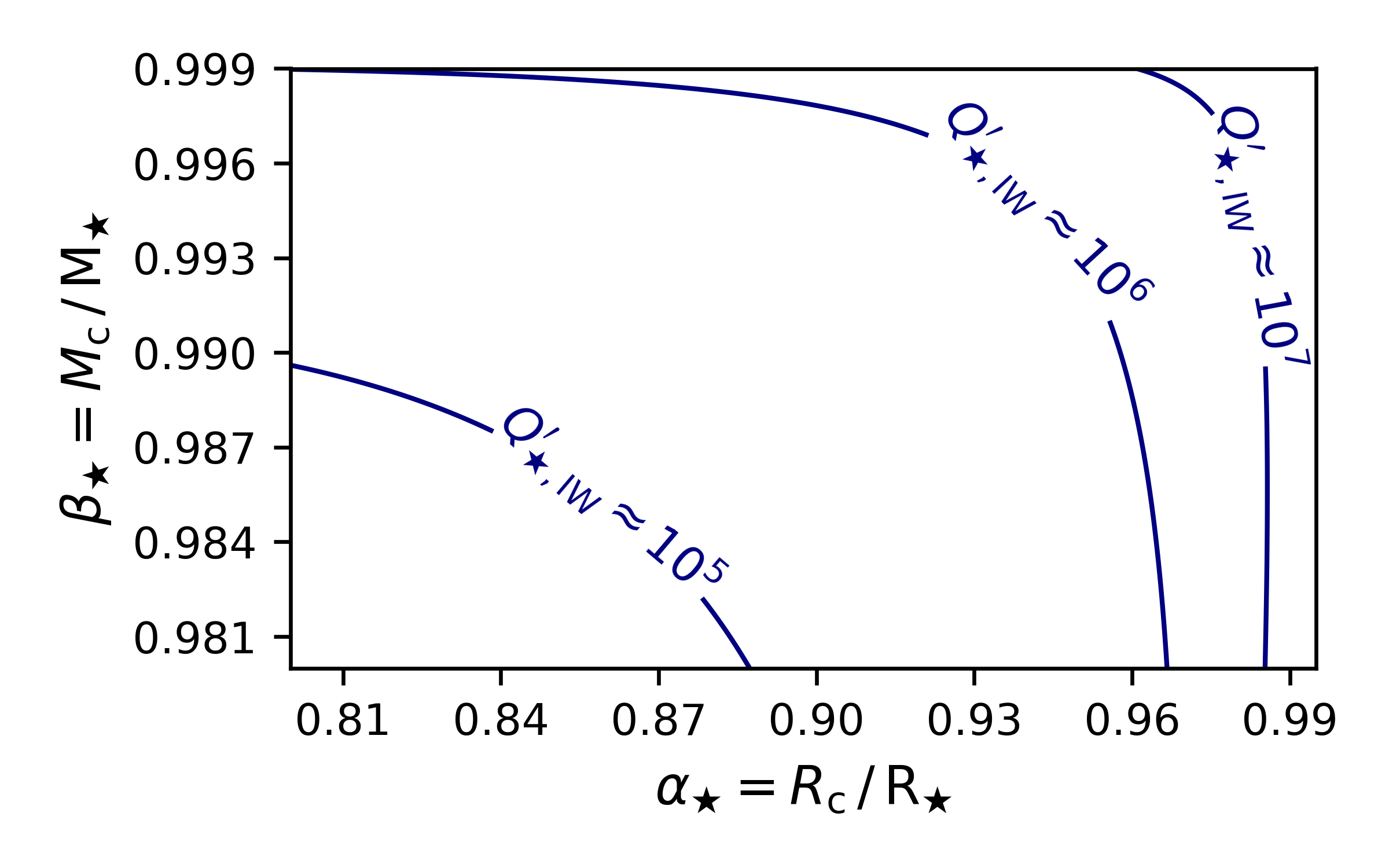}%
    \caption{Tidal quality factor $Q\sub{\star,IW}'$ as a function of the stellar aspect ratios $\alphas$ and $\betas$ of TOI-2109.}
    \label{fig:contourplot}
\end{figure}

TOI-2109 is a relatively young star, but the stellar rotational period ($\prot$), albeit very short, is longer than the planet's orbital period. This is a significant physical condition for the evolution of the system since both mechanisms (IWs and IGWs) act simultaneously while $\omega\leq2\Os$. For small eccentricities and rapidly rotating stars, this holds true only if $\porb\geq\frac{\prot}{2}$. Figures~\ref{fig:tidaldecay_young} and~\ref{fig:tidaldecay_old} show that a part of the dynamical evolution of TOI-2109b is dominated by the dissipation of IWs. However, once the planet reaches a semi-major axis where $\porb\ngeq\frac{\prot}{2}$ (i.e. very short orbital periods), IWs in the stellar convective envelope were no longer excited and IGWs were the only active tidal dissipation mechanism for the rest of the planet's orbital evolution. IGWs in the stellar radiative regions were modeled following \citet{Barker2020}.

\begin{table*}
\setlength{\tabcolsep}{3pt}
\centering
\caption{System parameters.}
\begin{tabular}{lcccccccc}
\hline\hline
System & $T_0$ (BJD$_\mathrm{TDB}$) & $P_0$ (d) & $\Mp (\Mjup)$ & $\Rp (\Rjup)$ & $\Mstar$ ($\Msun$) & $\Rstar$ ($\Rsun$) & $\prot$ (d) & $\tage$ (Gyr)\\
\hline
TOI-2109 & $2459378.459370^{+0.000059}_{-0.000059}$ & $0.67247414^{+0.00000028}_{-0.00000028}$ & $5.02^{+0.75}_{-0.75}$ & $1.347^{+0.047}_{-0.047}$ & $1.447^{+0.075}_{-0.078}$ & $1.698^{+0.062}_{-0.057}$ & $1.05^{+0.04}_{-0.04}$ & $1.77^{+0.88}_{-0.68}$\\
\bottomrule
\end{tabular}
\label{tab:usp}
\begin{minipage}{\textwidth}
\vspace{0.2cm}
\footnotesize {\bf Note}. All the parameters shown were extracted from \citet{Wong2021}.
\end{minipage}
\end{table*}

The planetary interior aspect ratios ($\alphap$ and $\betap$) were set according to \citet{Guenel2014}. Also, we assumed that TOI-2109b's core is solid \citep{Guenel2014} and has a rigidity of $4.46\times10^{10}$ Pa, a value calculated from tidal interactions of Jupiter with Io \citep{Lainey2012}. Despite it has been recently shown that Jupiter's core might be diluted due to an innermost region rich in heavy elements \citep[e.g.,][]{Muller2020}, such an assumption simplifies the estimation of the viscoelastic dissipation of TOI-2109b's core using Maxwell's linear approximation. For TOI-2109, Figure~\ref{fig:contourplot} shows the expected aspect ratios $\alphas$ and $\betas$ described in subsection~\ref{sec:k2q_IW} and embedded in the tidal evolution equations (Section~\ref{sec:eqs}). These values agree with those reported by \citet{Gallet2017} for a star of $\Mstar\approx1.45\,\Msun$, producing $Q\sub{\star,\,IW}'$ between $10^5$ and $10^7$ \citep{Barker2020}.

We studied the orbital evolution of TOI-2109b using a numerical integrator that alternates between non-stiff and stiff methods according to the dynamics of the system. To calculate the instantaneous change of the quantities in
Equations~(\ref{eq:dnpdt})--(\ref{eq:dosdt}), we assumed that the planet's orbital plane is co-planar to the normal plane of the stellar spin vector (i.e. the sky-projected obliquity $\lambda$). This is very close to the actual $\lambda$ measured in \citet{Wong2021}, which revealed a well-aligned orbit with $\lambda=1.7^\circ\pm1.7^\circ$. We used the system parameters reported by \citet{Wong2021} and summarized in Table~\ref{tab:usp}.

We modeled the orbital evolution of TOI-2109b until its arrival at the Roche limit, where it would be destroyed by strong stellar tides \citep{Roche1849}. Different authors have shown that the evolution of USP-Js strongly depends on the mechanism through which the host star dissipates internal energy \citep[e.g.,][]{Alvarado2021}. This energy is a result of the tidal interactions with the planetary companion, driving the exchange of angular momentum between the stellar spin and the orbit of the planet, which can shrink the planetary orbit and change the sky-projected orbital obliquity.

We studied the system for $\prot/{\rm sin} i_*=1.05$ d, as derived by \citet{Wong2021} using spectroscopic data. When modeling the orbital evolution of TOI-2109b, the results for different values within the uncertainty range of the stellar rotation period were negligible and, therefore, we only present the results from simulations for the nominal value of $\prot$. As previously mentioned, the dominance of each dissipation mechanism considered here (see Section~\ref{sec:dissmech}) will strongly depend on the stellar age as described next.

\subsection{`Young' TOI-2109}
\label{sec:youngstar}

Given uncertainties in the age of TOI-2109, the lower-end and central value of the stellar age, $\tage$ (i.e., 1.09 and 1.77 Gyr, see Table~\ref{tab:usp}) do not comply with the requirements to damp IGWs in radiation zones; that is, the planetary companion would have to be much more massive than TOI-2109b, thus indicating that IGWs are not in the wave breaking regime and are instead linear. Our results imply that tidal dissipation due to IGWs in a `young' TOI-2109 will be likely orders of magnitude less efficient at $Q\sub{\star,\,IGW}'\approx10^{8.5}$ \citep[cf.][]{Barker2020}. That said, most of the dissipation then resulted from the excitation of IWs in the stellar convective envelope, for which we studied the forward orbital evolution of TOI-2109b using the model described in subsection~\ref{sec:k2q_IW}.

\begin{figure}
    \centering
    \hspace{-0.5cm}
    \includegraphics[scale=0.62]{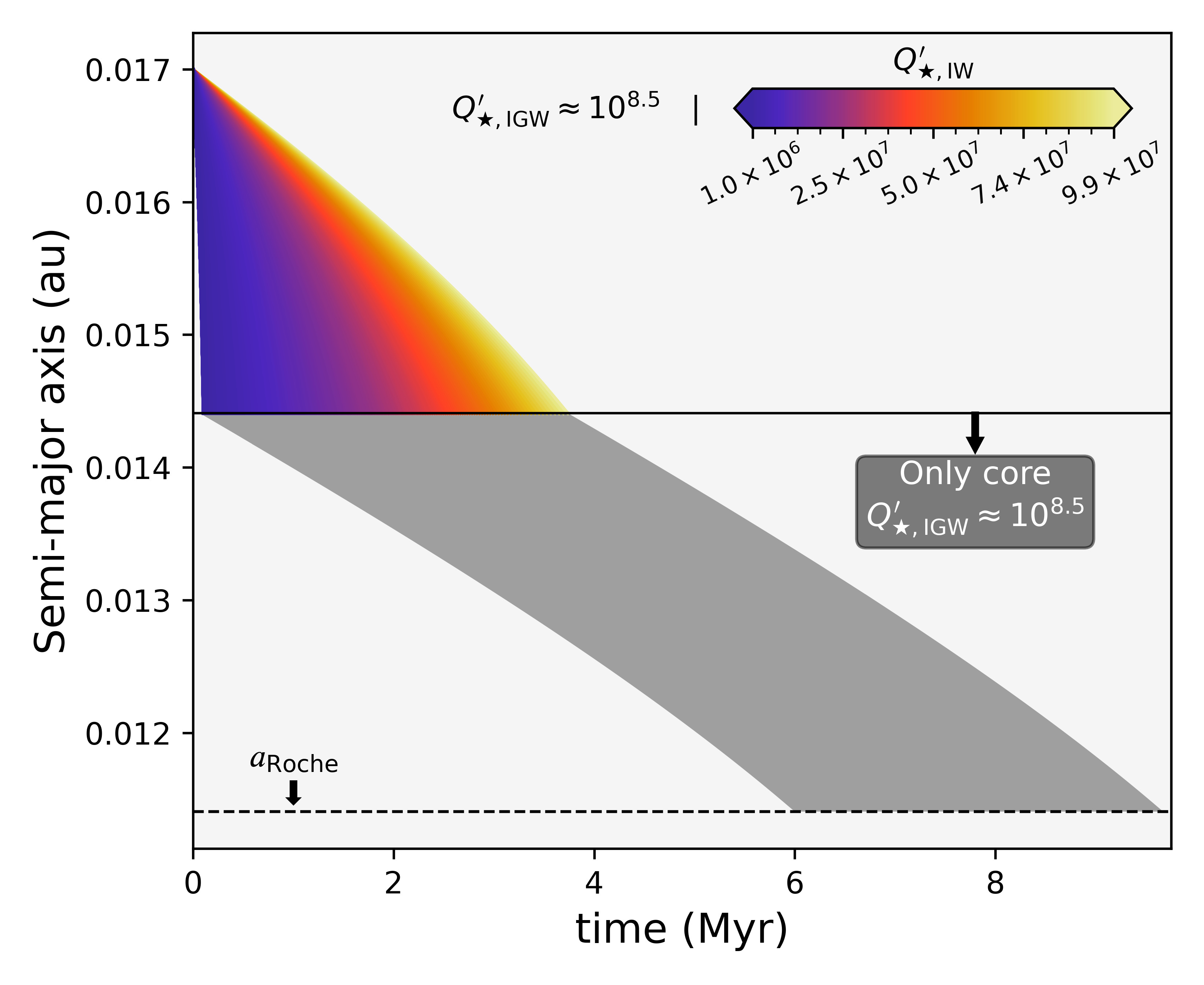}%
    \caption{{\it Top:} Semi-major axis of TOI-2109b as a function of time for a stellar age 1.09--1.77 Gyr. The colors stand for  different $Q\sub{\star,\,IW}'$ in the evolutionary equations (Section~\ref{sec:eqs}) and follow the formalism of Section~\ref{sec:dissmech}, while $Q\sub{\star,\,IGW}'\approx10^{8.5}$ for a `young' host star.
    \label{fig:tidaldecay_young}}
\end{figure}

Figure~\ref{fig:tidaldecay_young} shows the evolution of the semi-major axis for different values of $Q\sub{\star,\,IW}'$, where tidal dissipation arises mostly from excited IWs. From these tidal simulations we calculate the decay rate of the orbital period as $\dpdtide{-4.186}{0.797}$ at $3\sigma$. Using the formulation of \citet{Goldreich1966} and $Q_\star' = 3Q_\star/(2k_{2, \star})$, this leads to 95\% confidence lower limit of the decay timescale, $\tau=\frac{P}{\dot{P}}$, $\tau\sub{TIDE}\gtrsim6$ Myr, and $95\%$ confidence lower limit of \mbox{$Q\sub{\star,\,TIDE}'>2.3\times10^{7}$}. The latter is in agreement with the expected values of a type-F star with $1.445\,\Msun$ \citep{Barker2020}, as well as with the expected aspect ratios $\alphas$ and $\betas$ (Figure~\ref{fig:contourplot}) in the current scenario of stellar tidal dissipation for a `young' TOI-2109 \citep[cf.][]{Gallet2017}.

It is particularly interesting that, despite this system is an ideal candidate to study orbital decay in USP Jupiters, the aforementioned decay rate for TOI-2109b might be an indicator of a constant orbital period. Furthermore, a direct comparison of the decay rate calculated here to those predicted by \citet{Wong2021} (i.e., 10--740\,$\msyr$), indicates that the lower end might still hold while the upper end can be ruled out. As shown later in Section~\ref{sec:tt_analysis}, this is further supported by observational constraints using space-based \footnote{This includes {\it TESS} observations from three sectors (see Table~\ref{tab:TESSdata}).} and ground-based data of TOI-2109.

\subsection{`Old' TOI-2109}
\label{sec:oldstar}

As mentioned before, the dissipation of IGWs is driven by the combination of stellar age, stellar mass, and planetary mass. These, together, control the damping of such waves in the stellar radiative regions. In general, the damping of IGWs in young stars requires extremely large planetary masses for IGWs to be in the wave breaking regime. This is why in the previous section the excitation of IGWs was significantly inefficient and most of the tidal dissipation was driven by IWs in the stellar convective envelope. However, it is also worth noting that all stars with $\Mstar>0.9\,\Msun$ will present a steady decrease in $M_\mathrm{crit}$ before reaching 10 GYr \citep{Barker2020}. This means that for a given stellar mass, IGWs might be fully damped in a later stage of stellar evolution as $M_\mathrm{crit}$ decreases to lower values.

\begin{figure}
    \centering
    \hspace{-0.5cm}
    \includegraphics[scale=0.62]{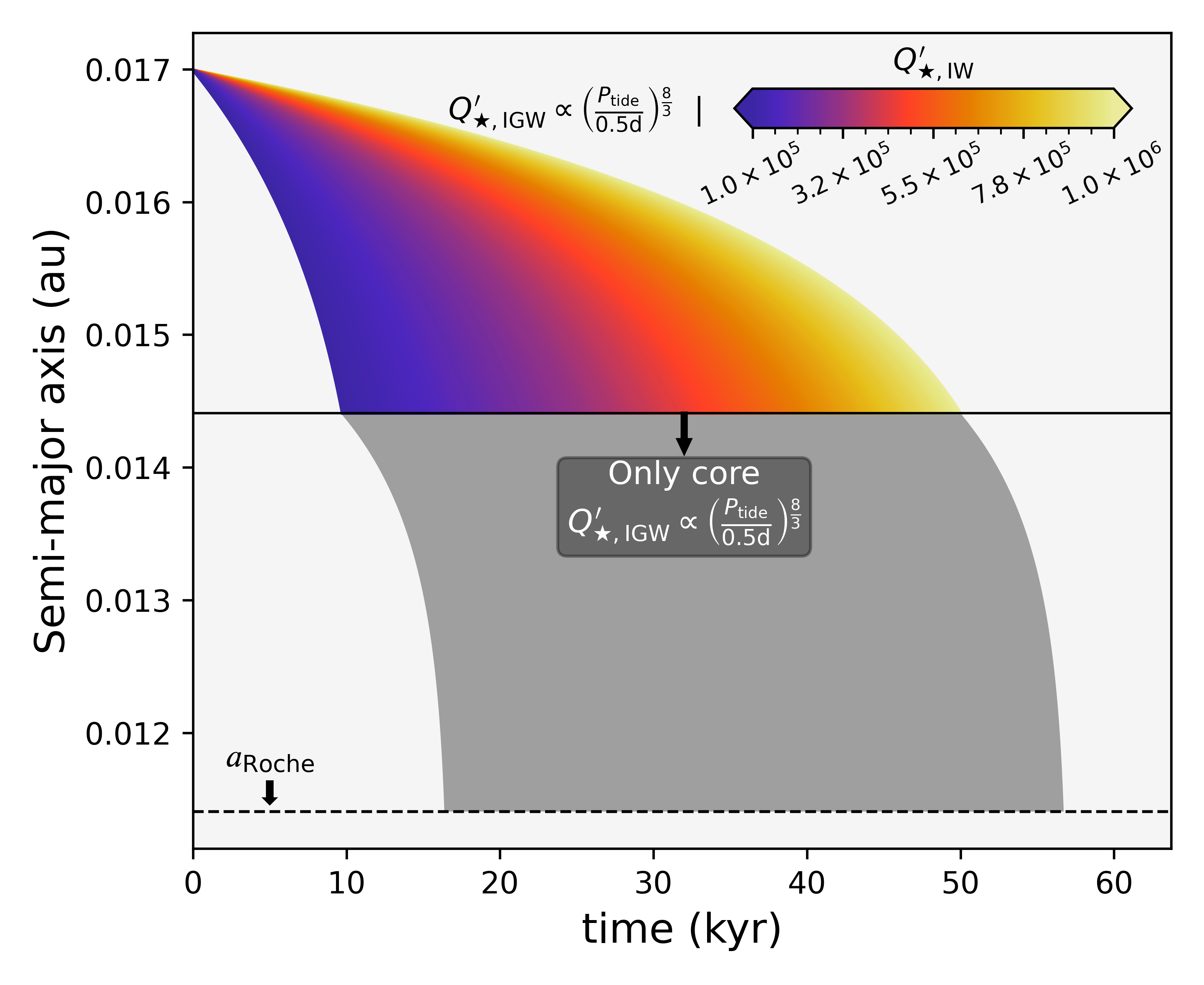}%
    \caption{Same as Figure~\ref{fig:tidaldecay_young} but for $\tage=2.65$ Gyr.}
    \label{fig:tidaldecay_old}
\end{figure}

In light of the above, we notice that if the stellar age of TOI-2109 is closer to the upper-end of the 1-sigma range (i.e., $\tage=2.65$ Gyr, see Table~\ref{tab:usp}), IGWs might break in the stellar radiation region as $M\sub{crit}$ would be between 1 and 10 Jupiter masses. Following \citet{Barker2020}, we notice that a planetary mass of $\sim5\,\Mjup$ would make IGWs that are launched from the radiative/convective regions of the star go into the wave breaking regime and subsequently be fully damped. An `old' TOI-2109 would follow $Q\sub{\star,\, IGW}'\propto(P_\mathrm{tide}/0.5\mathrm{d})^{\frac{8}{3}}$, presenting a more efficient tidal dissipation due to IGWs. In Figure~\ref{fig:tidaldecay_old}, we show the semi-major axis as a function of time and $Q\sub{\star}'$. The latter is now a significant combination of $Q\sub{\star,\,IW}'$ and $Q\sub{\star,\,IGW}'$, noting that during half of the orbital evolution both dissipation mechanisms (IWs and IGWs) acted together.

Once $\porb\ngeq\frac{\prot}{2}$ (i.e., no excitation of IWs), IGWs acted alone and dominated the angular momentum transport in the star. This behavior was expected because IGWs, if efficiently damped, produce the strongest tidal dissipation rates for short orbital periods \citep{Barker2020}. From the tidal simulations presented in Figure~\ref{fig:tidaldecay_old}, we calculated the orbital decay rate as \mbox{$\dpdtide{-1107}{648}$}, resulting in a decay timescale $\tau\sub{TIDE}\gtrsim17$ kyr for the $Q\sub{\star}'$ that were studied. The previous decay rate puts a lower limit of $Q\sub{\star,\,TIDE}>8.7\times10^4$, which for the adopted values of $\alphas$ and $\betas$ is close to the lowest-limit of $Q\sub{\star}'$ in F-stars, ranging from $10^5$ to $10^7$ \citep{Ogilvie2007,Lanza2011}.

Distinct values of $C$ in $Q\sub{\star,\, IGW}'=C\,(P_\mathrm{tide}/0.5\mathrm{d})^{\frac{8}{3}}$ could change the previous lower limits, which corresponded to a \mbox{$C=3$}. \citet{Barker2020} found that $C$ can vary from 1 to 3 for main-sequence stars with 0.5--1.1$\,\Msun$, thus expecting \mbox{$C\gtrsim3$} for TOI-2109. To quantify these changes, we increased $C$ by an {\it inflated factor} of two (as $C$ would just be slightly larger), which led to a decrease of $\sim34\%$ in $\dot{P}\sub{TIDE}$, and an increase of $\sim45\%$ in $\tau\sub{TIDE}$ and $Q\sub{\star,\,TIDE}'$. As studied in the next Section, however, the results from this scenario---regardless of $C$---are still vastly inconsistent with the constraints from transit-timing observations reported to date. Hence, seeking conservative values in this study case, we kept $C=3$ in all our simulations for an `old' TOI-2109.

In comparison to USP-J systems such as NGTS-10b and WASP-19b that could spin up the rotational rate of their host star by about 30 per cent \citep[see][]{Alvarado2021}, we found that the rotational rate of TOI-2109 was not significantly spun up during the shrinking of the orbit of its planetary companion. This may be a consequence of the moment of inertia of TOI-2109 (i.e. $\Is\propto\Mstar\Rstar^2$), which is significantly higher than that of the previously mentioned systems. All our dynamical simulations resulted in final stellar rotation periods longer than the saturation rotation period of TOI-2109 (i.e. $\bar{P}\sub{\star}=0.559$ d), using the saturation rate $\Obar$ provided by \citet{Cameron1994}. We conclude that during the orbital evolution of TOI-2109b, the host star's magnetic activity remains a function of the stellar rotational rate and a constant stellar activity level is not achieved.

\begin{figure*}
    \centering
    \includegraphics[scale=0.35]{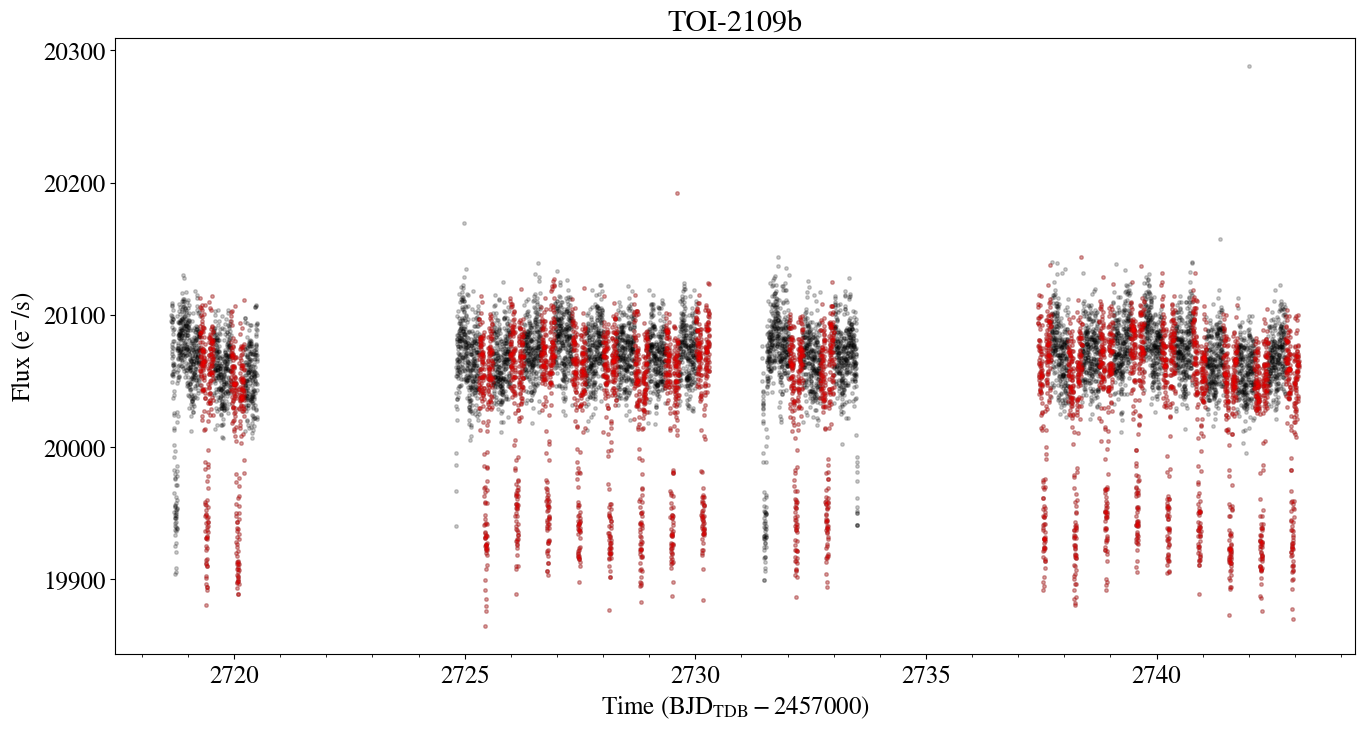}
    \caption{{\it TESS} light curve of TOI-2109b. The groups of red points are the transit intervals, which are centered on the predicted transit times and extend for four transit durations. {\it TESS} data from Sector 52 with a two-minute cadence.}
    \label{fig:timeseries}
\end{figure*}

\begin{figure*}
    \centering
    \includegraphics[scale=0.32]{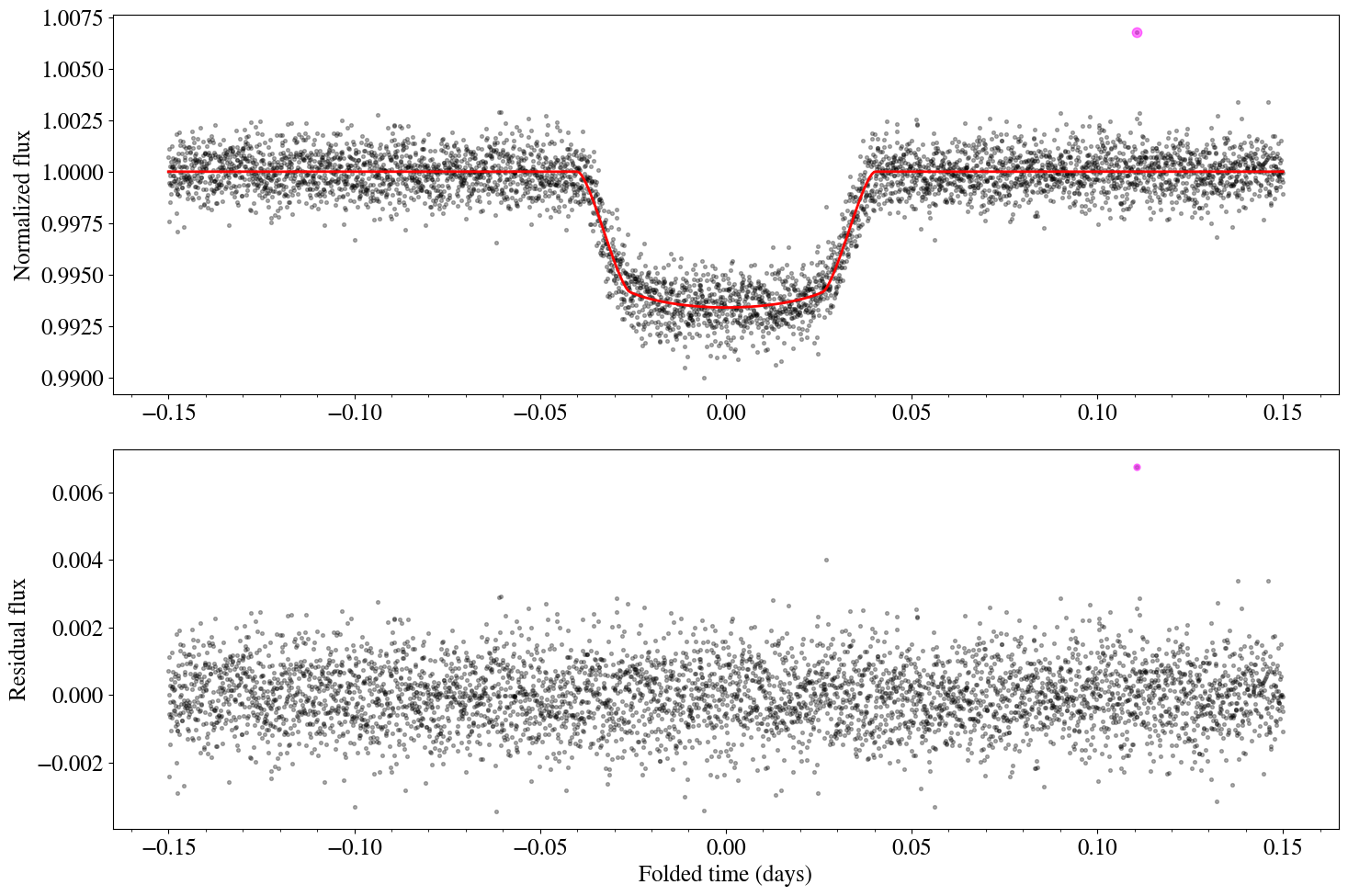}
    \caption{Folded light curve of TOI-2109b using the data of Figure~\ref{fig:timeseries}.}
    \label{fig:folded_lightcurve}
\end{figure*}

\begin{figure*}
    \centering
    \includegraphics[scale=0.3]{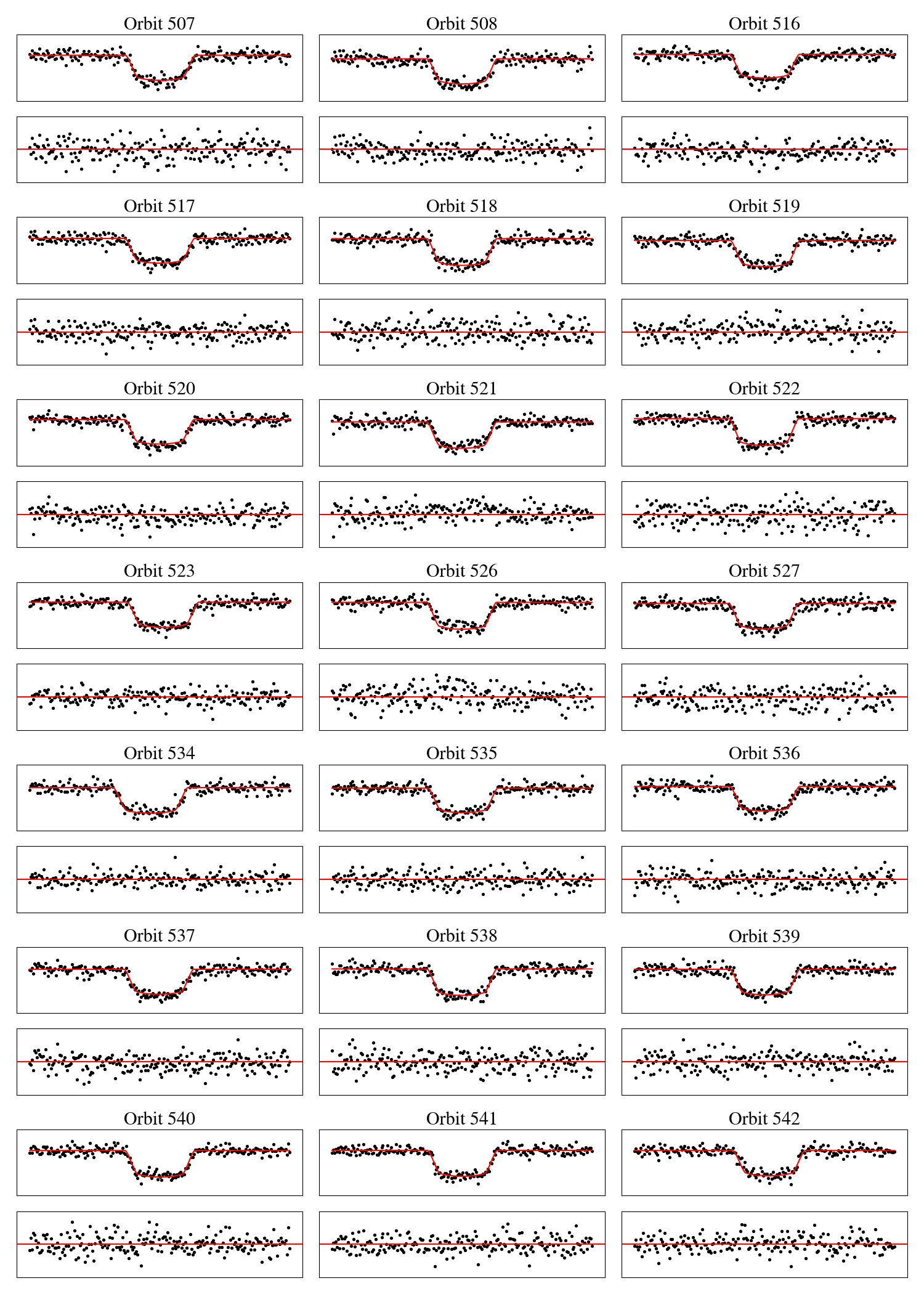}
    \caption{Individual transits extracted from Figure~\ref{fig:timeseries}. Red curves represent best-fitting models.}
    \label{fig:individualtransits_1800}
\end{figure*}





\begin{table}
    \centering
    \caption{{\it TESS} Data Processed for TOI-2109b}
    \begin{tabular}{llcccccl}
         \hline
         \hline
         &Data Source&  &  Cadence (s)&  &{\it TESS} Sector& &Year\\\hline
         &TESS-SPOC&  &  1800&  & 25& &2020\\
         &SPOC&  &  120&  & 52& &2022\\
         &SPOC&  &  20&  & 79& &2024\\
         \hline
    \end{tabular}
    \label{tab:TESSdata}
\end{table}

\section{Transit-timing analysis}
\label{sec:tt_analysis}

In this section, we use space-based and ground-based long-baseline data of TOI-2109 to place new observational constraints on the decay rate of the planetary companion as well as on the stellar tidal quality factor $Q\sub{\star}'$. The existing data for TOI-2109 and how we used it to extract each mid-transit time will be explained in the following two subsections.

\subsection{TESS and CHEOPS Data}
\label{sec:tessdata}

TOI-2109 or TIC 392476080, as listed in the {\it TESS} Input Catalog \citep[TIC,][]{Stassun2018} has been observed by {\it TESS} in: 2020 (May 13 to June 8, Sector 25), 2022 (May 18 to June 13, Sector 52), and 2024 (May 21 to Jun 18, Sector 79). All {\it TESS} data was downloaded from the Mikulski Archive for Space Telescopes (MAST\footnote{\href{https://mast.stsci.edu/}{https://mast.stsci.edu/}}) using Lightkurve, a Python package for {\it Kepler} and {\it TESS} data analysis \citep{Lightkurve2018}. In Table~\ref{tab:TESSdata} we list the {\it TESS} data we processed in this work, indicating the source, cadence, and sector. The simple aperture photometry (SAP) light-curve files\footnote{These consist of the flux extracted from the optimized aperture on the full-frame images (FFI).} were  previously extracted via the Science Processing and Operations Center (SPOC) as part of the {\it TESS} Light Curves from Full Frame Images (TESS-SPOC) High Level Science Products project \citep{Caldwell2020}.

We process the SAP light curves ourselves and measure the mid-transit time of each observation, using the formalism and code for transit-timing analysis presented in \citet{Ivshina2022}---hereafter IV22---which has been employed to calculate transit times in several other systems. We input the ephemeris of TOI-2109b from \citet{Wong2021} into the transit-timing code of IV22\footnote{This code was originally designed for data with a nominal cadence of 2, 10, and 30 minutes. The formalism, however, can be easily adapted to be used with a cadence of 20 seconds (i.e., for the data from sector 79).}, which analyses individual transit intervals by only selecting transit data with optimal time coverage and removing any transit interval with a lack of data points. Occasional outliers were removed by following the Median Absolute Deviation (MAD) of residuals used in IV22, which is also employed in all the photometric data to identify and reject transits of unusually poor quality. We present the flux time series of TOI-2109 from {\it TESS} Sector 52, obtained using IV22 and with all selected transits for transit-timing analysis marked in red (Figure~\ref{fig:timeseries}).

To remove photometric trends such as stellar variability, IV22 fits a model to the out-of-transit data and uses it to normalize ("rectify") the entire light curve. The individual transits are then phase-folded (Figure~\ref{fig:folded_lightcurve}) to extract key parameters that describe the geometry of the transits, including $R\sub{p}/R\sub{\star}$, $a/R_{\star}$, impact parameter ($b$), and limb-darkening law. Once reliable values for these parameters were obtained from the best-fit transit model of the phase-folded light curve, they were used to fit each individual transit and determine the corresponding mid-transit time. In Figure~\ref{fig:individualtransits_1800}, we show the selected transits of TOI-2109b for Sector 52 along with the best-fit transit model (in red) and residuals. Each subplot in this figure is titled with the number of the orbit, as calculated from the zero epoch in \citet{Wong2021}. To estimate the mid-transit times (and uncertainties) of the selected transits, we used a Markov Chain Monte Carlo (MCMC) method. We applied each step of the previous process\footnote{For a more detailed explanation, see section 3 ({\it TESS} Data Analysis) in \citet{Ivshina2022}.} to the {\it TESS} data in Table~\ref{tab:TESSdata}, measuring 31, 21, and 22 reference mid-transit times for Sectors 25, 52, and 79, respectively. These will be refined later in subsection~\ref{sec:ttvs_midtt} for transit-timing analysis.

From 2021 to 2023, TOI-2109 was observed 31 times by CHEOPS, 26 of which contain transits with significant coverage. The data is fully accessible, for example, via the CHEOPS archive\footnote{\href{https://cheops-archive.astro.unige.ch/archive_browser/}{https://cheops-archive.astro.unige.ch/archive\_browser/}} through the Data and Analysis Center for Exoplanets (DACE) website\footnote{\href{https://dace.unige.ch/cheopsDatabase/?}{https://dace.unige.ch/cheopsDatabase/?}} of the University of Geneva. However, we did not process the data ourselves but instead adopted the results of \citet{Harre2024}\footnote{Available at \href{https://cdsarc.cds.unistra.fr/viz-bin/cat/J/A+A/692/A254}{https://cdsarc.cds.unistra.fr/viz-bin/cat/J/A+A/692/A254}}, who calculated the 26 reference mid-transit times in the CHEOPS data. They extracted these values from their best-fit model after preparing and reducing the data with two different photometry pipelines: the CHEOPS Data Reduction Pipeline \citep[{\texttt DRP};][]{Hoyer2020} and the PSF Imagette Photometric Extraction \citep[{\texttt PIPE}; see e.g.,][and references therein]{Brandeker2022}. The resulting  mid-transit values from both photometry pipelines are systematically within 1-sigma of each other's uncertainties. Since both sets of values are not substantially different from each other, we chose to work with the {\texttt DRP} mid-transit values \citep[see table C.1 in][]{Harre2024}.

\begin{figure*}
    \centering
    \includegraphics[width=\textwidth]{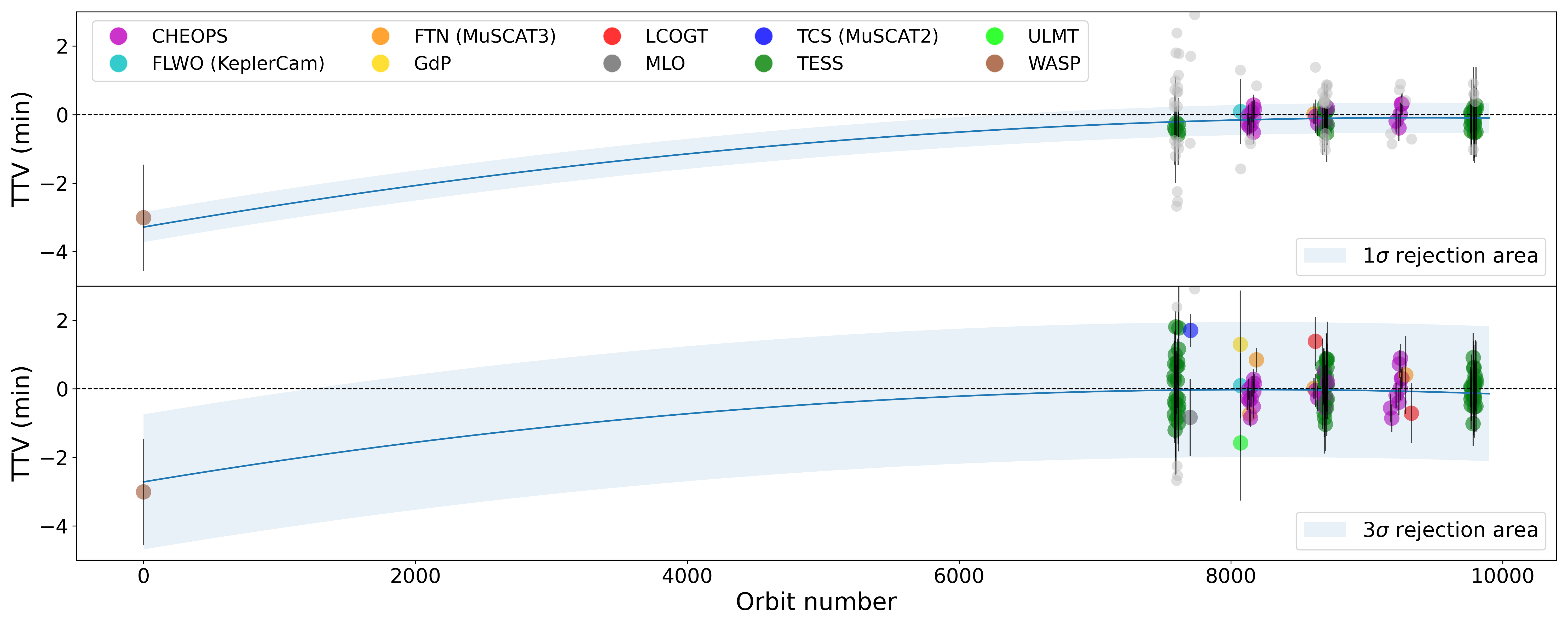}
    \caption{TTVs of TOI-2109b calculated with {\texttt PdotQuest}, from an MCMC fit using a quadratic model. The fit is performed for a 1-sigma (upper panel) and 3-sigma (lower panel) rejection areas. (i.e., the blue shaded regions).}
    \label{fig:quadratic}
\end{figure*}

\begin{table}
    \centering
    \caption{Ground-based Data Processed for TOI-2109b}
    \begin{tabular}{cccc}
         \hline
         \hline
         Telescope&  Cadence (s)&  Filter& Year\\\hline
         FLWO KeplerCam&  5& $z'$& 2020, 2021\\
         ULMT&  16 – 32& $r',\,i'$& 2020, 2021\\
         TCS MuSCAT2& 3& $g',\,r',\,i',\,z\sub{s}$& 2020\\
         MLO&  30& $I$& 2020\\
         LCOGT& 8 – 40& $g',\,B,\,i',\,z\sub{s}$& 2020 – 2023\\
         WBRO&  100& $R$& 2021\\
         GdP&  20& $i'$& 2021\\
         FTN MuSCAT3& 12 – 33& $g',\,r',\,i',\,z\sub{s}$& 2021\\
         \hline
    \end{tabular}
    \label{tab:GROUNDdata}
\end{table}

\begin{table*}
\centering
\caption{Ephemeris of TOI-2109b}
\begin{tabular}{ccccccccc}
\hline
\hline
System & $N_\mathrm{transits}$ & $T_0$ (BJD$_\mathrm{TDB}$) & Uncertainty (d) & $P_0$ (d) & Uncertainty (d) & $\dot{P}\sub{OBSV}$ (ms yr ${ }^{-1}$) & $\sigma$ Rejection & Reference \\
\hline TOI-2109b & 53 & 2453885.6884 & 0.0009 & 0.67247457 & 0.00000022 & $-2.344 \pm 1.260$ & 1 & This work\\
TOI-2109b & 102& 2453885.6888 & 0.0009 & 0.67247455 & 0.00000022 & $-2.616 \pm 1.285$ & 3 & This work\\
\hline
\end{tabular}
\label{tab:ephemerides}
\begin{minipage}{\textwidth}
\vspace{0.2cm}
\footnotesize {\bf Note}. The table contains ephemeris for TOI-2109b fitted by a quadratic model from potential TTVs. The ephemeris were fitted using two different sigma rejection schemes. This table is available in its entirety in machine-readable form.
\end{minipage}
\end{table*}

\subsection{Ground-Based Data}
\label{sec:grounddata}

Long-term variations in the orbital period of TOI-2109b can only be detected by having transit data from a well-sampled, long-time baseline. As mentioned before, ever since TOI-2109b was discovered back in 2020, {\it TESS} has only observed it three times. Therefore, to have a four-year baseline that is better sampled to do transit-timing analysis, we use archival data obtained from The Exoplanet Follow-up Observing Program (\href{https://exofop.ipac.caltech.edu/tess/}{ExoFOP}). These include the 20 partial-transit observations of \citet{Wong2021} collected by multiple ground-based telescopes: the Fred Lawrence Whipple Observatory (FLWO, 1.2 m) on Mt. Hopkins, Arizona, USA; the University of Louisville Manner Telescope (ULMT, 0.61 m) at Mt. Lemmon, Arizona, USA; Las Cumbres Observatory Global Telescope network \citep[LCOGT, 0.4–1.0 m;][]{Brown2013}; the Maury Lewin Astronomical Observatory (MLO, 0.36 m) in California, USA; the Telescopio Carlos Sánchez (TCS, 1.5 m) at Teide Observatory, Spain; the Wild Boar Remote Observatory (WBRO, 0.24 m) near Florence, Italy; the Grand-Pra (GdP, 0.4 m) Observatory in Switzerland; and the Faulkes Telescope North (FTN, 2.0 m) at Haleakala Observatory on Maui, Hawai’i. The ground-based data of TOI-2109 covers from 2020 to 2023, with a cadence as low as 3 s and using different filters, as shown in Table~\ref{tab:GROUNDdata}.

We independently analyzed each dataset at their corresponding epoch, using the PyTransit package \citep{Parviainen2015} for the light-curve modeling. For each dataset, a \citet{Mandel2002} model was fitted to the phase-folded light curve. The parameters that we used for this modeling included planet-to-star radius ratio $\Rp/\Rstar$, stellar density in g\,cm$^{-3}$, impact parameter $b$, orbital period $P$, and zero epoch $T_0$. To fit the eccentricity and argument of periastron, PyTransit used the parametrization $\sqrt{e}\,{\rm cos} \omega$ and $\sqrt{e}\,{\rm sin} \omega$. We followed a similar process to that of subsection~\ref{sec:tessdata}, that is, we fitted first the phase-folded light curve to extract reliable values of the parameters mentioned before and then use these values to fit each individual transit, thus recovering its reference mid-transit time. We fully process all the ground-based data ourselves, recovering 20 reference mid-transit times.

Data from the All-Sky Automated Survey for Supernovae (ASAS-SN) were available for TOI-2109, which spans for over 10 years. However, these observations have significant variations in cadence and most of them do not even complete a partial transit. For the purpose of calculating mid-transit times, at least a partial transit is needed to identify the mid-point and thus recover the transit time. For this reason, we decided to neglect ASAS-SN data and only use data from other dedicated follow-up missions, as previously mentioned. Moreover, SuperWASP-N \citep[Wide Angle Search for Planets;][]{Pollacco2006} observed TOI-2109 in several fields, from 2006 to 2011. We do not process this data ourselves but instead use the only reference mid-transit time recovered by \citet{Harre2024}, as most of the data was filtered out due to poor transit-timing quality caused by bad transit coverage and/or high scatter. This data point, however, is significant and provides a longer baseline that can allow us to further constrain the orbital decay rate of TOI-2109b.

\subsection{TTVs and mid-transit times to constrain \texorpdfstring{$\dot{P}$}{P-dot}}
\label{sec:ttvs_midtt}

Transit-timing variations (TTVs) can be used as possible indicators of orbital decay. TTVs can be recovered by comparing the reference mid-transit times with calculated, refined values that are obtained from a fitting process to the transit-timing data. For TOI-2109b, we use the reference mid-transit times previously found (subsections~\ref{sec:tessdata} and~\ref{sec:grounddata}) into a quadratic ephemeris model described as follows

\beq{eq:Tn}
T_N=T_{N-1}+P_{N-1}
\eeq
\beq{eq:Pn}
P_N=P_{N-1}+\dot{P}(P_{N-1}+P_N)/2
\eeq
where $T_0$, $P_0$, $N$, and $T_N$ is the reference mid-transit time, the initial orbital period of the planet, the number of orbits since the reference transit, and the 
calculated mid-transit time, respectively. The constant rate of change of the orbital period, $\dot{P}$, in Equation~(\ref{eq:Pn}) implies that there is a small difference between the orbital period of two successive transits. In this model, we fit the free parameters $T_0$, $P_0$, and $\dot{P}$, taking initial guesses for $T_0$ and $P_0$ from \citet{Wong2021}.

To fit the previous model to our reference transit-timing data\footnote{This can be found at \href{https://github.com/JAAlvarado-Montes/TOI-2109b}{https://github.com/JAAlvarado-Montes/TOI-2109b}}, we use the software package {\texttt PdotQuest} provided by \citet{Wang2024}. By inputting the reference mid-transit times of TOI-2109b and their uncertainties into {\texttt PdotQuest}, we calculated each new mid-transit time and associated uncertainty, thus recovering $\dot{P}$ in the process. Internally, {\texttt PdotQuest} uses the expression $\frac{{\rm d}P}{{\rm d}t}=\frac{1}{P}\frac{{\rm d}P}{{\rm d}N}$ to adopt the quadratic form of equations~(\ref{eq:Tn}) and~(\ref{eq:Pn}) as in \citet{Maciejewski2021}.

It is worth noting that, despite {\it TESS} provides the majority of mid-transit times, the photometric quality of {\it TESS} data hinders the transit-timing analysis. An extreme example of this is the data from {\it TESS} Sector 25 where some of the reference mid-transit times had uncertainties up to $\sim$ 10 min. For a better fit to the transit-timing data, any mid-transit points with $\Delta T_0\gtrsim2$ min were clipped out before using {\texttt PdotQuest}. This allowed us to use a wider variety of data sources while only keeping the data with a higher  timing precision for a more robust fit. This process filtered out 8 mid-transit times from the ground-based data and 6 from {\it TESS} data (5 from sector 25 and 1 from sector 79). To avoid including data with high scatter, {\texttt PdotQuest} uses an iterative data-clipping technique to remove outliers in the TTVs \citep[see section 3 in][]{Wang2024}, also called $n$-sigma rejection areas. Such outliers are represented as gray dots in Figure~\ref{fig:quadratic}, which shows TTVs of TOI-2109b for a $1\sigma$ (upper panel) and $3\sigma$ (lower panel) rejection area, respectively. While the $3\sigma$ fit used 102 mid-transit times out of the 117 that were provided---including several high-scatter data points---the results from this fit are completely consistent with the $1\sigma$ calculation that only uses the 53 more precise mid-transit times.

\begin{figure*}
    \centering
    \includegraphics[width=\textwidth]{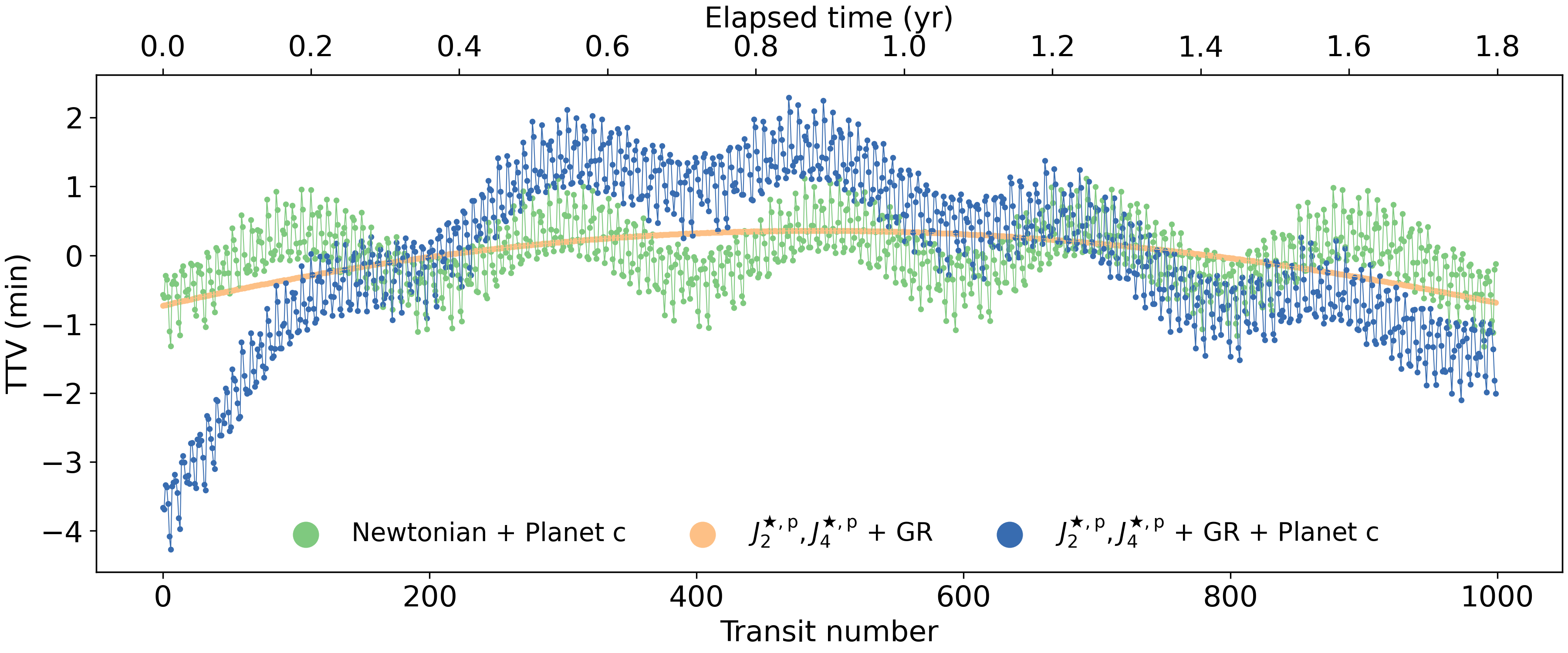}
       \caption{TTVs of TOI-2109b under different dynamical scenarios. The green points represent TTVs resulting from Newtonian gravitational interactions with an additional planetary companion (planet c, $P_c \approx 1.125\,\mathrm{d}$, $e_c=0.05$). The orange points depict the secular TTV trend as a result of rotationally induced oblateness (characterized by quadrupole $J_2$ and octupole $J_4$) of both the star and the planet, along with general relativistic precession effects. Finally, the blue points incorporate all these effects simultaneously, demonstrating the superposition of both short-term perturbations from the planetary companion and long-term secular drifts from oblateness and relativistic corrections.}
    \label{fig:ttvnomig}
\end{figure*}

The complete results from our transit-timing analysis are reported in Table~\ref{tab:ephemerides}. We can see how, for both sigma-rejection areas, our calculated orbital decay rate is only compatible with the lower limit of the range 10--740\,$\msyr$ first reported by \citet{Wong2021}, ruling out any $\dot{P}\gtrsim10\,\msyr$. Also, \citet{Harre2024} calculated an orbital decay rate of just $\dot{P}\sim-1\,\msyr$ by using ground- and space-based data. Our best-fitting value of $\dot{P}$ is consistent with these results and is based on one additional sector of {\it TESS} data (i.e., sector 79) that was unavailable in previous works \citep{Wong2021,Harre2024}. With a a cadence of 20 seconds, this dataset provided a well-mapped transit timeseries with an improved precision on the mid-transit times as compared to those from previous {\it TESS} sectors (i.e., 25 and 52). These 22 new data points from {\it TESS} sector 79 were included in the TTV analysis of TOI-2109b and allowed us to further constrain its orbital decay rate, yielding a value of $\dpdt{-2.616}{1.285}$ at $3\sigma$. This allowed us to put a lower limit $Q\sub{\star,\,OBSV}'>3.7\times10^7$, which is in agreement with those of F-stars that range from $10^5$ to $10^7$.

\subsection{TTVs for a non migrating planet}
\label{sec:ttv_nomig}

TTVs in a compact, ultra-short-period planetary system such as TOI-2109b can arise from multiple physical mechanisms, including gravitational interactions with additional planetary companions \citep{Harre2024}, general relativistic effects, and deviations from spherical symmetry (oblateness) of both the planet and the host star \citep[see, e.g.,][and references therein]{Sucerquia2025}.
To quantify these effects, we performed numerical simulations using the \texttt{REBOUNDx} package \citep{rebound,reboundx}, systematically evaluating each mechanism's contribution to the TTV signal. First, we simulated gravitational perturbations induced by a proposed planetary companion with a mass of $0.2 \Mjup$, a period of \( P_{\mathrm{c}} \sim 1.125 \) days, and a moderate eccentricity (\( e = 0.05 \)). For TOI-2109b, we adopted \( e = 0.04 \), consistent with the parameter space explored by \citet{Harre2024}. The resulting TTVs exhibit a clear, short-term periodic signature with amplitudes of approximately 1--2 minutes, emphasizing the companion's significant gravitational influence. Importantly, this model does not include any tidal orbital decay; instead, it illustrates how precession and dynamical perturbations alone can generate a quasi-secular TTV signal that may superficially resemble decay, highlighting the need for caution in its interpretation.

To evaluate the impact of rotationally induced oblateness, we calculated the gravitational quadrupole moment \( J_2 \) for TOI-2109b assuming synchronous rotation (rotational period equal to orbital period, \(P_{\mathrm{rot}} = P_{\mathrm{orb}}\)). This yielded \( J_2 \approx 3.119 \times 10^{-4}\), indicative of significant oblateness comparable to Saturn. Additionally, we included stellar oblateness characterized by a similar quadrupole moment calculation (assuming a stellar rotation period of about 1.14 days and a Love number \(k_2=0.03\)), and general relativistic corrections appropriate for the system's compactness and eccentricity.

\begin{equation}
\label{eq:j2star}
J_\mathrm{2,\,p} = \frac{3}{4}\frac{\Omega_\mathrm{p}^2 R_{\mathrm{p}}^3}{G M_{\mathrm{p}}}, \quad \text{with } \Omega_\mathrm{p} = \frac{2\pi}{P_{\mathrm{rot}}},
\end{equation}

We performed these simulations using the fast Wisdom-Holman symplectic integrator \texttt{WHFast} \citep{wh,reboundwhfast}, with a timestep of \(1\times10^{-5}\) years. This integrator is particularly suited for compact systems, offering high numerical stability and accuracy over long integration times. Transit times in our simulations were precisely determined by detecting intersections of the planet with the \(y=0\) plane for \(x>0\), refined with a bisection algorithm to achieve a temporal accuracy of \(10^{-8}\) years.

The resulting simulated transit times were analyzed by removing a linear trend through least-squares fitting, thus isolating the variations attributable to the studied dynamical mechanisms. All physical effects were consistently incorporated via \texttt{REBOUNDx}'s specialized modules. General relativity was included using the \texttt{gr\_potential} module, while the oblateness of both planet and star were implemented using the \texttt{gravitational\_harmonics} module. For the planet, synchronous rotation dictated the above-mentioned \(J_2\) and \(J_4 = -J_2/5\) values.

To account for the stellar quadrupole moment, we adopted a rotational period of 1.4 days and a stellar Love number \( k_2 = 0.03 \). The corresponding \( J_2^\star \) was computed following

\begin{equation}
\label{eq:j2plan}
J_{2,\,\star} = \frac{k_2 \, \Omega_\star^2 \, R_\star^3}{3 G M_\star}, \quad \text{where} \quad \Omega_\star = \frac{2\pi}{P_{\mathrm{rot},\star}},
\end{equation}
and the octupole moment was set as \( J_4^\star = -J_2^\star/5 \), consistent with the quadrupolar potential expansion. This value captures the contribution of stellar oblateness to the long-term orbital precession of TOI-2109b.

Although Equations~(\ref{eq:j2star}) and~(\ref{eq:j2plan}) share a similar functional form—both scaling as \( \Omega^2 R^3 / (G M) \)—they originate from different physical considerations. Equation~(\ref{eq:j2star}) is derived from the hydrostatic equilibrium of a rotating fluid planet with a simplified internal structure (e.g., a polytrope of index \(n=1\)), yielding a direct expression for \(J_2\) without requiring additional structural parameters \citep[e.g.,][]{Murray2000}. In contrast, Equation~(\ref{eq:j2plan}) incorporates the stellar Love number \(k_2\), which quantifies the star’s response to rotational deformation and encapsulates details of its internal density distribution and stratification \citep[e.g.,][]{Kopal1960,Zharkov1978}. The use of \(k_2\) makes the stellar expression more general and applicable to objects with complex internal profiles. Despite their formal similarity, the coefficients and underlying assumptions of each equation reflect the differing physical regimes of gaseous giant planets and rotating stars.

Figure~\ref{fig:ttvnomig} shows the resulting TTVs under different dynamical scenarios. The green dots represent TTVs caused solely by gravitational interactions with the additional planetary companion. The orange dots incorporate both stellar and planetary oblateness effects along with general relativistic precession, showing a slower, secular trend with smaller amplitudes. Finally, the blue dots integrate all these mechanisms, revealing a signal composed of both periodic, short-term variations from the planetary companion and longer-term secular drift from oblateness and relativistic effects.

Our results reveal distinct dynamical signatures in the TTV signal: short-term variability is primarily driven by the gravitational perturbations of an external planetary companion, while stellar and planetary oblateness, together with general relativistic precession, induce subtle but coherent secular trends. This decomposition is essential for accurately interpreting observed TTVs and disentangling the underlying physical mechanisms shaping the orbital architecture of compact systems like TOI-2109b—particularly in the context of long-duration, high-precision monitoring.

\section{Discussion}
\label{sec:disc}

TOI-2109b is one of the best candidates for orbital decay among known ultra-short-period Jupiters (USP-Js), standing out due to its remarkably short orbital period and the characteristics of its host star. Specifically, TOI-2109 exhibits the highest rotational rate as well as the largest size and mass among the twelve known USP-J hosts. All these features may have profound implications for the fate of TOI-2109b, which we analyze by considering the tidal dissipation via inertial waves (IWs) in the convective envelope of the planet and the star, as well as internal gravity waves (IGWs) in the radiative regions of the star.

Rare processes, such as core-envelope decoupling in a star, could lead to a partial exchange of angular momentum with its companion and result in rotation at a different rate (and perhaps even with a different orientation). This could play an important role in the observable signatures of orbital decay (e.g., mid-transit times), and some authors have included these effects in simple tidal models \citep[e.g.,][]{Barker2009,Winn2010,Alvarado-Montes2022}. However, such processes are still poorly constrained and pose a challenge for the general validity of core-envelope decoupling models over stellar evolutionaryS timescales. Internal mechanisms of angular momentum transport---such as gravity waves, magnetic fields, or magnetohydrodynamic turbulence---would be expected to suppress strong differential rotation over stellar (and tidal) evolutionary timescales, making the approach used in this work effective for studying the tidal evolution of USP-Js overall.

In the event of orbital decay, the rate at which the period may decrease is determined by the efficiency of the stellar tidal processes that drive the exchange of angular momentum with the planetary orbit. For an F-type star, these processes are expected to be insufficient to trigger a significant exchange of angular momentum that could eventually lead to orbital decay. However, it is also worth noting that large uncertainties in the stellar age imply that such processes are poorly constrained and so is their efficiency. Within the framework of this study, the main difference between our two study cases (i.e., a `young' and `old' host star; Section~\ref{sec:orbdecay}) is the efficiency at which TOI-2109 dissipates IGWs in its radiative regions. 

Our study reveals that an `old' host star would induce an extremely fast decay of the orbital period of TOI-2109b, at a rate of $\dpdtide{-1107}{648}$ for \mbox{$Q\sub{\star,\,TIDE}'>8.7\times10^4$}. The upper limit of this orbital decay rate would favor the upper limit of $10-740\,\msyr$ found by \citet{Wong2021}. Such a fast decay rate implies very efficient tidal processes that further intensify as the planet decays in its orbit since $Q\sub{\star,\, IGW}'\propto(P_\mathrm{tide}/0.5\mathrm{d})^{\frac{8}{3}}$. The results of this scenario, however, should be taken with a grain of salt. Such values would produce observable changes in the orbital period and reference mid-transit time of the order of seconds and hours, respectively \citep{Cameron2018}. These changes should have already occurred in the 4+ years since the initial discovery of TOI-2109b and, given their calculated magnitude, the baseline established by {\it TESS} data should be sufficient to confirm them. However, no noticeable changes are detected in the data from {\it TESS} sectors 52 and 79, where our transit-timing analysis revealed mid-transit times that are inconsistent with a rapidly decaying orbit (Section~\ref{sec:tt_analysis}).

Conversely, the planetary companion is significantly more stable against orbital decay if the host star TOI-2109 is younger. In this case, orbital decay could potentially occur at a rate of $\dpdtide{-4.186}{0.797}$, which entails $Q\sub{\star,\,TIDE}'>2.3\times10^{7}$ and is more consistent with a constant-period orbit. This slow orbital decay rate is in agreement with the lower limit of the range 10--740\,$\msyr$ \citep{Wong2021} but slightly larger than $\dot{P}\sim-1\,\msyr$ reported by \citet{Harre2024}. A low orbital decay rate can be evidence of inefficient tidal processes that could result from how fast the host star rotates (i.e., $\prot/{\rm sin} i_*\approx1$ d) and how this dampens the exchange of angular momentum from the planetary orbit to the stellar spin. For an `old' host star, where TOI-2109b undergoes fast orbital decay, such a damping may be exceeded by: 1) the more efficient dissipation of IWs due to a large stellar age \citep{Barker2020}; 2) an enhanced IGW-dissipation associated to a decreasing orbital period (i.e., a decaying orbit).

It is worth pointing out that the values of $Q\sub{\star}'$ for the `young' and `old' scenarios of TOI-2109 are also consistent with other tidal dissipation processes. For example,  
\citet{Weinberg2024} studied orbital decay of hot Jupiters due to weakly nonlinear tidal dissipation. Despite a precise comparison to TOI-2109 is difficult owing to a tighter parameter space grid in their work, a rough interpolation between the 0.5 and 1.0-day orbital periods shows that an `old' TOI-2109 could still produce $Q\sub{\star}'\sim10^5$ and $\tau\sim10^4$\,yr. This suggests that, even if TOI-2109 is not strictly in the wave-breaking regime (perhaps it is a bit too young), tidal dissipation can still be significant if $\tage\approx2.5$\,Gyr. On the contrary, if $\tage\lesssim2$\,Gyr, their results show that weakly nonlinear dissipation is much less efficient and $\tau\sim10^7$\,yr, which is consistent with the young-star scenario considered in this work.

The use of transit-timing variations (TTVs), in a similar fashion to that of WASP-12b \citep{Yee2020} and Kepler-1658b \citep{Vissapragada2022}, can be a valuable method to further study the possibility of orbital decay in this system. Our TTV analysis of TOI-2109b also seems to favor a rather constant-period orbit (Section~\ref{sec:tt_analysis}), with a best-fitting decay rate of just $\dpdt{-2.616}{1.285}$ and a corresponding $Q\sub{\star,\,OBSV}'>3.7\times10^7$. This is consistent with the aforementioned tidal scenario of a `young' TOI-2109 where $\dpdtide{-4.186}{0.797}$, supporting too the lower limit of 10--740\,$\msyr$ \citep{Wong2021} and $\dot{P}\sim-1\,\msyr$ \citep{Harre2024}.

\begin{figure}
    \centering
    \hspace{-0.5cm}
    \includegraphics[scale=0.62]{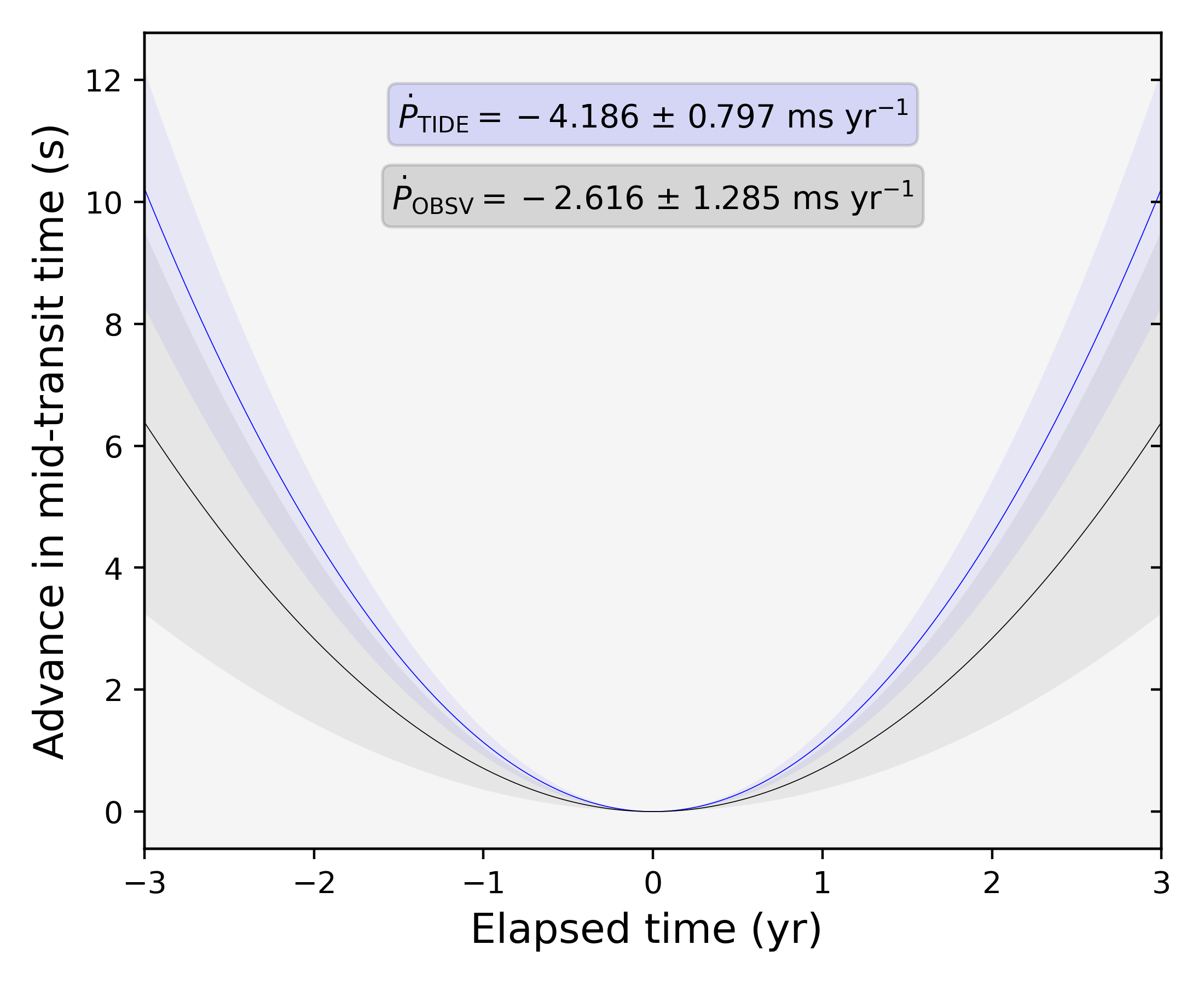}
    \caption{Advance in mid-transit times of TOI-2109b measured in human timescales. This is for the scenario of a `young' host star.}
    \label{fig:human_deltas}
\end{figure}

The lower bound of $\pobsv$ may be interpreted as tentative evidence of orbital decay, although at such a slow rate that the orbital period would appear constant even in long baselines. Still, orbital decay can slowly and progressively produce changes in expected mid-transit times. Both results presented here for TOI-2109b, namely, $\ptide$ and $\pobsv$ would produce an advance in mid-transit time of the order of a few seconds ($\Delta T_0\sim10^{-4}$ d) for a 3-year baseline (see Figure~\ref{fig:human_deltas}), which could be detected with high-cadence observations. The dependence of $\ptide$ on the mechanisms of energy dissipation and its correlation with $\pobsv$ implies that its verification (or falsification) can provide valuable insights into the validity of tidal models.

For this system, however, large uncertainties of the mid-transit times represent a significant problem for TTV modeling as for any predictions on orbital decay. Future photometric and spectroscopic data will therefore be necessary to further constrain such a scenario. To {\it definitely} confirm or deny any significant TTVs is linked to the necessity of understanding USP planets at a deeper level. This stresses the need of collecting higher-quality observations so that an improved transit-timing analysis can be carried out. For instance, higher-cadence frames with {\it TESS} may offer an optimal precision for mid-transit times. Also, a continuous follow-up with CHEOPS, which provides a timing precision of just a few seconds \citep{Harre2024}, could be ideal for further investigation of TOI-2109b. Precise transit-timing can also be done with observations from the Hubble Space Telescope (HST) or James Webb Space Telescope (JWST). Low-time use of these (and other) instruments is possible due to the tight orbit of TOI-2109b, also allowing us to characterize its atmosphere \citep[see, e.g.,][]{Cont2025}.

In this system, the existence of a yet-to-be-found close outer companion may be feasible \citep{Harre2024}. If confirmed, this could be a factor that contributes to making tidal processes inefficient: the low values of the orbital decay rates \(\ptide\) and \(\pobsv\) may further support this scenario. Moreover, the presence of an external perturber can make the orbit of TOI-2109b become slightly eccentric ($e>0.003$), and the sinusoidal variations of apsidal precession would also be more significant. In fact, for short precession periods, TTVs produced by apsidal precession may look identical to those of orbital decay.

For a complete and accurate interpretation of the TTVs in TOI-2109b, especially in the context of potential orbital decay, it is essential to first account for all other dynamical sources that may contribute to the TTV signal. To quantify the magnitude of general relativistic effects, stellar and planetary oblateness, and the gravitational perturbation of a potential outer companion \citep{Harre2024}, a continuous follow-up of this system will be necessary---as well as to study their distinct amplitudes (subsection~\ref{sec:ttv_nomig}). These must be understood and disentangled prior to attributing any remaining trends to orbital migration. New observational campaigns aimed at collecting high-quality timing data of TOI-2109b will help us identify TTVs and  unlock the ability to study how they may result from other effects. Our layered modeling approach allows us to isolate the contributions that could be ascribed to tidal evolution.

It is also worth noting that many current studies of ultra-short-period planets do not explicitly incorporate the dynamical effects arising from the non-sphericity of the star or the planet when interpreting TTV signals. In systems where tidal locking and rapid rotation are expected, these effects can introduce coherent secular trends that may be wrongly attributed to orbital decay or apsidal precession if not properly modeled. We highlight the importance of including such contributions in TTV analyses to avoid misinterpretations and to build a more comprehensive understanding of the complex orbital dynamics at play in USP systems.



From temporary hot Jupiters that only exist for short periods \citep{Stephan2018}, to hot Jupiters that paradoxically host water in their atmospheres \citep{Parmentier2018}, USP-Js represent perhaps the most extreme planetary systems and thus the insights we gain from studying these systems are likely to have far-reaching implications. For example, recent infrared observations conducted by \citet{De2023} have unveiled significant details pertaining to a likely USP-J being engulfed by its star. This could shed light on the chemical composition and migration patterns of USP-Js, helping us understand the accretion of rocky material that may result in selective enrichment of refractory elements and lithium on stellar surfaces \citep{Oh2018}. These discoveries offer valuable insights into the internal mass distribution and dynamics of USP-Js, underscoring the critical importance of investigating their tidal evolution.

\section*{Data Availability}
All the {\it TESS} light-curve files of TOI-2109b used in this paper can be found in MAST: \dataset[10.17909/jaks-bm57]{http://dx.doi.org/10.17909/jaks-bm57} (sector 25),
\dataset[10.17909/j0sa-9a12]{http://dx.doi.org/10.17909/j0sa-9a12} (sector 52), and \dataset[10.17909/fb38-0214]{http://dx.doi.org/10.17909/fb38-0214} (sector 79). All the timing data for TTV analysis can be directly downloaded from \href{https://github.com/JAAlvarado-Montes/TOI-2109b}{https://github.com/JAAlvarado-Montes/TOI-2109b}.

\section*{Acknowledgements}
The authors thank the anonymous referees whose comments and insights significantly improved the quality of this work. Thanks to Dr Anita Hafner for valuable suggestions on the manuscript. JAA-M acknowledges funding from the Macquarie University Research Fellowship (MQRF). This project was supported by the European Research Council (ERC) under the European Union Horizon Europe research and innovation program (grant agreement No. 101042275, project Stellar-MADE). JIZ is supported by Minciencias Grant 1115-852-70719/70939. This research made use of the NASA's Astrophysics Data System (ADS) and NASA Exoplanet Archive, operated by the California Institute of Technology under NASA's Exoplanet Exploration Program.




\begin{thebibliography}{}
\expandafter\ifx\csname natexlab\endcsname\relax\def\natexlab#1{#1}\fi
\providecommand{\url}[1]{\href{#1}{#1}}
\providecommand{\dodoi}[1]{doi:~\href{http://doi.org/#1}{\nolinkurl{#1}}}
\providecommand{\doeprint}[1]{\href{http://ascl.net/#1}{\nolinkurl{http://ascl.net/#1}}}
\providecommand{\doarXiv}[1]{\href{https://arxiv.org/abs/#1}{\nolinkurl{https://arxiv.org/abs/#1}}}

\bibitem[{E.~R. {Adams} {et~al.}(2024){Adams}, {Jackson}, {Sickafoose}, {Morgenthaler}, {Worters}, {Stubbers}, {Carlson}, {Bhure}, {Dekeyser}, {Huang}, \& {Weinberg}}]{Adams2024}
{Adams}, E.~R., {Jackson}, B., {Sickafoose}, A.~A., {et~al.} 2024, \bibinfo{title}{{Doomed Worlds. I. No New Evidence for Orbital Decay in a Long-term Survey of 43 Ultrahot Jupiters},} \psj, 5, 163, \dodoi{10.3847/PSJ/ad3e80}

\bibitem[{B. {Akinsanmi} {et~al.}(2019){Akinsanmi}, {Barros}, {Santos}, {Correia}, {Maxted}, {Bou{\'e}}, \& {Laskar}}]{Akinsanmi2019}
{Akinsanmi}, B., {Barros}, S.~C.~C., {Santos}, N.~C., {et~al.} 2019, \bibinfo{title}{{Detectability of shape deformation in short-period exoplanets},} \aap, 621, A117, \dodoi{10.1051/0004-6361/201834215}

\bibitem[{B. {Akinsanmi} {et~al.}(2024){Akinsanmi}, {Barros}, {Lendl}, {Carone}, {Cubillos}, {Bekkelien}, {Fortier}, {Flor{\'e}n}, {Collier Cameron}, {Bou{\'e}}, {Bruno}, {Demory}, {Brandeker}, {Sousa}, {Wilson}, {Deline}, {Bonfanti}, {Scandariato}, {Hooton}, {Correia}, {Demangeon}, {Smith}, {Singh}, {Alibert}, {Alonso}, {Asquier}, {B{\'a}rczy}, {Barrado Navascues}, {Baumjohann}, {Beck}, {Beck}, {Benz}, {Billot}, {Bonfils}, {Borsato}, {Broeg}, {Buder}, {Charnoz}, {Csizmadia}, {Davies}, {Deleuil}, {Delrez}, {Ehrenreich}, {Erikson}, {Farinato}, {Fossati}, {Fridlund}, {Gandolfi}, {Gillon}, {G{\"u}del}, {G{\"u}nther}, {Heitzmann}, {Helling}, {Hoyer}, {Isaak}, {Kiss}, {Lam}, {Laskar}, {Lecavelier des Etangs}, {Magrin}, {Maxted}, {Mecina}, {Mordasini}, {Nascimbeni}, {Olofsson}, {Ottensamer}, {Pagano}, {Pall{\'e}}, {Peter}, {Piazza}, {Piotto}, {Pollacco}, {Queloz}, {Ragazzoni}, {Rando}, {Rauer}, {Ribas}, {Santos}, {S{\'e}gransan}, {Simon}, {Stalport}, {Szab{\'o}}, {Thomas}, {Udry}, {Van Grootel}, {Venturini},
  {Villaver}, \& {Walton}}]{Akinsanmi2024}
{Akinsanmi}, B., {Barros}, S.~C.~C., {Lendl}, M., {et~al.} 2024, \bibinfo{title}{{The tidal deformation and atmosphere of WASP-12 b from its phase curve},} \aap, 685, A63, \dodoi{10.1051/0004-6361/202348502}

\bibitem[{M.~E. {Alexander}(1973){Alexander}}]{Alexander1973}
{Alexander}, M.~E. 1973, \bibinfo{title}{{The Weak Friction Approximation and Tidal Evolution in Close Binary Systems},} \apss, 23, 459, \dodoi{10.1007/BF00645172}

\bibitem[{E. {Alvarado} {et~al.}(2024){Alvarado}, {Bostow}, {Patra}, {Jacobus}, {Baer-Way}, {Jennings}, {Pichay}, {deGraw}, {Vidal}, {Chander}, {Altunin}, {Brendel}, {Ehrich}, {Sunseri}, {May}, {Punjabi}, {Gendreau-Distler}, {Risin}, {Brink}, {Zheng}, \& {Filippenko}}]{AlvaradoEF2024}
{Alvarado}, E., {Bostow}, K.~B., {Patra}, K.~C., {et~al.} 2024, \bibinfo{title}{{Searching for tidal orbital decay in hot Jupiters},} \mnras, 534, 800, \dodoi{10.1093/mnras/stae2062}

\bibitem[{J.~A. {Alvarado-Montes}(2022){Alvarado-Montes}}]{Alvarado-Montes2022}
{Alvarado-Montes}, J.~A. 2022, \bibinfo{title}{{Tidally induced migration of TESS gas giants orbiting M dwarfs},} \mnras, 517, 2831, \dodoi{10.1093/mnras/stac2741}

\bibitem[{J.~A. {Alvarado-Montes} \& C. {Garc{\'\i}a-Carmona}(2019){Alvarado-Montes} \& {Garc{\'\i}a-Carmona}}]{Alvarado2019}
{Alvarado-Montes}, J.~A., \& {Garc{\'\i}a-Carmona}, C. 2019, \bibinfo{title}{{Orbital decay of short-period gas giants under evolving tides},} \mnras, 486, 3963, \dodoi{10.1093/mnras/stz1081}

\bibitem[{J.~A. {Alvarado-Montes} {et~al.}(2021){Alvarado-Montes}, {Sucerquia}, {Garc{\'\i}a-Carmona}, {Zuluaga}, {Spitler}, \& {Schwab}}]{Alvarado2021}
{Alvarado-Montes}, J.~A., {Sucerquia}, M., {Garc{\'\i}a-Carmona}, C., {et~al.} 2021, \bibinfo{title}{{The impact of tidal friction evolution on the orbital decay of ultra-short-period planets},} \mnras, 506, 2247, \dodoi{10.1093/mnras/stab1081}

\bibitem[{A. {Bailey} \& J. {Goodman}(2019){Bailey} \& {Goodman}}]{Bailey2019}
{Bailey}, A., \& {Goodman}, J. 2019, \bibinfo{title}{{Understanding WASP-12b},} \mnras, 482, 1872, \dodoi{10.1093/mnras/sty2805}

\bibitem[{A.~J. {Barker}(2011){Barker}}]{Barker2011}
{Barker}, A.~J. 2011, \bibinfo{title}{{Three-dimensional simulations of internal wave breaking and the fate of planets around solar-type stars},} \mnras, 414, 1365, \dodoi{10.1111/j.1365-2966.2011.18468.x}

\bibitem[{A.~J. {Barker}(2016){Barker}}]{Barker2016}
{Barker}, A.~J. 2016, \bibinfo{title}{{Non-linear tides in a homogeneous rotating planet or star: global simulations of the elliptical instability},} \mnras, 459, 939, \dodoi{10.1093/mnras/stw702}

\bibitem[{A.~J. {Barker}(2020){Barker}}]{Barker2020}
{Barker}, A.~J. 2020, \bibinfo{title}{{Tidal dissipation in evolving low-mass and solar-type stars with predictions for planetary orbital decay},} \mnras, 498, 2270, \dodoi{10.1093/mnras/staa2405}

\bibitem[{A.~J. {Barker} {et~al.}(2024){Barker}, {Efroimsky}, {Makarov}, \& {Veras}}]{Barker2024}
{Barker}, A.~J., {Efroimsky}, M., {Makarov}, V.~V., \& {Veras}, D. 2024, \bibinfo{title}{{On the orbital decay of the gas giant Kepler-1658b},} \mnras, 527, 5131, \dodoi{10.1093/mnras/stad3530}

\bibitem[{A.~J. {Barker} \& G.~I. {Ogilvie}(2009){Barker} \& {Ogilvie}}]{Barker2009}
{Barker}, A.~J., \& {Ogilvie}, G.~I. 2009, \bibinfo{title}{{On the tidal evolution of Hot Jupiters on inclined orbits},} \mnras, 395, 2268, \dodoi{10.1111/j.1365-2966.2009.14694.x}

\bibitem[{A.~J. {Barker} \& G.~I. {Ogilvie}(2010){Barker} \& {Ogilvie}}]{Barker2010}
{Barker}, A.~J., \& {Ogilvie}, G.~I. 2010, \bibinfo{title}{{On internal wave breaking and tidal dissipation near the centre of a solar-type star},} \mnras, 404, 1849, \dodoi{10.1111/j.1365-2966.2010.16400.x}

\bibitem[{S.~C.~C. {Barros} {et~al.}(2022){Barros}, {Akinsanmi}, {Bou{\'e}}, {Smith}, {Laskar}, {Ulmer-Moll}, {Lillo-Box}, {Queloz}, {Cameron}, {Sousa}, {Ehrenreich}, {Hooton}, {Bruno}, {Demory}, {Correia}, {Demangeon}, {Wilson}, {Bonfanti}, {Hoyer}, {Alibert}, {Alonso}, {Escud{\'e}}, {Barbato}, {B{\'a}rczy}, {Barrado}, {Baumjohann}, {Beck}, {Beck}, {Benz}, {Bergomi}, {Billot}, {Bonfils}, {Bouchy}, {Brandeker}, {Broeg}, {Cabrera}, {Cessa}, {Charnoz}, {Damme}, {Davies}, {Deleuil}, {Deline}, {Delrez}, {Erikson}, {Fortier}, {Fossati}, {Fridlund}, {Gandolfi}, {Mu{\~n}oz}, {Gillon}, {G{\"u}del}, {Isaak}, {Heng}, {Kiss}, {des Etangs}, {Lendl}, {Lovis}, {Magrin}, {Nascimbeni}, {Maxted}, {Olofsson}, {Ottensamer}, {Pagano}, {Pall{\'e}}, {Parviainen}, {Peter}, {Piotto}, {Pollacco}, {Ragazzoni}, {Rando}, {Rauer}, {Ribas}, {Santos}, {Scandariato}, {S{\'e}gransan}, {Simon}, {Steller}, {Szab{\'o}}, {Thomas}, {Udry}, {Ulmer}, {Van Grootel}, \& {Walton}}]{Barros2022}
{Barros}, S.~C.~C., {Akinsanmi}, B., {Bou{\'e}}, G., {et~al.} 2022, \bibinfo{title}{{Detection of the tidal deformation of WASP-103b at 3 {\ensuremath{\sigma}} with CHEOPS},} \aap, 657, A52, \dodoi{10.1051/0004-6361/202142196}

\bibitem[{M. {Benbakoura} {et~al.}(2019){Benbakoura}, {R{\'e}ville}, {Brun}, {Le Poncin-Lafitte}, \& {Mathis}}]{Benbakoura2019}
{Benbakoura}, M., {R{\'e}ville}, V., {Brun}, A.~S., {Le Poncin-Lafitte}, C., \& {Mathis}, S. 2019, \bibinfo{title}{{Evolution of star-planet systems under magnetic braking and tidal interaction},} \aap, 621, A124, \dodoi{10.1051/0004-6361/201833314}

\bibitem[{S. {Biswas} {et~al.}(2024){Biswas}, {Bisht}, {Jiang}, {Sariya}, \& {Parthasarathy}}]{Biswas2024}
{Biswas}, S., {Bisht}, D., {Jiang}, I.-G., {Sariya}, D.~P., \& {Parthasarathy}, K. 2024, \bibinfo{title}{{Probing the Possible Causes of the Transit Timing Variation for TrES-2b in the TESS Era},} \aj, 168, 176, \dodoi{10.3847/1538-3881/ad6d66}

\bibitem[{E. {Bolmont} \& S. {Mathis}(2016){Bolmont} \& {Mathis}}]{Bolmont2016}
{Bolmont}, E., \& {Mathis}, S. 2016, \bibinfo{title}{{Effect of the rotation and tidal dissipation history of stars on the evolution of close-in planets},} Celestial Mechanics and Dynamical Astronomy, 126, 275, \dodoi{10.1007/s10569-016-9690-3}

\bibitem[{L.~G. {Bouma} {et~al.}(2019){Bouma}, {Winn}, {Baxter}, {Bhatti}, {Dai}, {Daylan}, {D{\'e}sert}, {Hill}, {Kane}, {Stassun}, {Villasenor}, {Ricker}, {Vanderspek}, {Latham}, {Seager}, {Jenkins}, {Berta-Thompson}, {Col{\'o}n}, {Fausnaugh}, {Glidden}, {Guerrero}, {Rodriguez}, {Twicken}, \& {Wohler}}]{Bouma2019}
{Bouma}, L.~G., {Winn}, J.~N., {Baxter}, C., {et~al.} 2019, \bibinfo{title}{{WASP-4b Arrived Early for the TESS Mission},} \aj, 157, 217, \dodoi{10.3847/1538-3881/ab189f}

\bibitem[{A. {Brandeker} {et~al.}(2022){Brandeker}, {Heng}, {Lendl}, {Patel}, {Morris}, {Broeg}, {Guterman}, {Beck}, {Maxted}, {Demangeon}, {Delrez}, {Demory}, {Kitzmann}, {Santos}, {Singh}, {Alibert}, {Alonso}, {Anglada}, {B{\'a}rczy}, {Barrado y Navascues}, {Barros}, {Baumjohann}, {Beck}, {Benz}, {Billot}, {Bonfils}, {Bruno}, {Cabrera}, {Charnoz}, {Collier Cameron}, {Corral van Damme}, {Csizmadia}, {Davies}, {Deleuil}, {Deline}, {Ehrenreich}, {Erikson}, {Farinato}, {Fortier}, {Fossati}, {Fridlund}, {Gandolfi}, {Gillon}, {G{\"u}del}, {Hoyer}, {Isaak}, {Kiss}, {Laskar}, {Lecavelier des Etangs}, {Lovis}, {Luntzer}, {Magrin}, {Nascimbeni}, {Olofsson}, {Ottensamer}, {Pagano}, {Pall{\'e}}, {Peter}, {Piotto}, {Pollacco}, {Queloz}, {Ragazzoni}, {Rando}, {Rauer}, {Ribas}, {Scandariato}, {S{\'e}gransan}, {Simon}, {Smith}, {Sousa}, {Steller}, {Szab{\'o}}, {Thomas}, {Udry}, {Van Grootel}, {Walton}, \& {Wolter}}]{Brandeker2022}
{Brandeker}, A., {Heng}, K., {Lendl}, M., {et~al.} 2022, \bibinfo{title}{{CHEOPS geometric albedo of the hot Jupiter HD 209458 b},} \aap, 659, L4, \dodoi{10.1051/0004-6361/202243082}

\bibitem[{D.~J.~A. {Brown} {et~al.}(2011){Brown}, {Collier Cameron}, {Hall}, {Hebb}, \& {Smalley}}]{Brown2011}
{Brown}, D.~J.~A., {Collier Cameron}, A., {Hall}, C., {Hebb}, L., \& {Smalley}, B. 2011, \bibinfo{title}{{Are falling planets spinning up their host stars?},} \mnras, 415, 605, \dodoi{10.1111/j.1365-2966.2011.18729.x}

\bibitem[{T.~M. {Brown} {et~al.}(2013){Brown}, {Baliber}, {Bianco}, {Bowman}, {Burleson}, {Conway}, {Crellin}, {Depagne}, {De Vera}, {Dilday}, {Dragomir}, {Dubberley}, {Eastman}, {Elphick}, {Falarski}, {Foale}, {Ford}, {Fulton}, {Garza}, {Gomez}, {Graham}, {Greene}, {Haldeman}, {Hawkins}, {Haworth}, {Haynes}, {Hidas}, {Hjelstrom}, {Howell}, {Hygelund}, {Lister}, {Lobdill}, {Martinez}, {Mullins}, {Norbury}, {Parrent}, {Paulson}, {Petry}, {Pickles}, {Posner}, {Rosing}, {Ross}, {Sand}, {Saunders}, {Shobbrook}, {Shporer}, {Street}, {Thomas}, {Tsapras}, {Tufts}, {Valenti}, {Vander Horst}, {Walker}, {White}, \& {Willis}}]{Brown2013}
{Brown}, T.~M., {Baliber}, N., {Bianco}, F.~B., {et~al.} 2013, \bibinfo{title}{{Las Cumbres Observatory Global Telescope Network},} \pasp, 125, 1031, \dodoi{10.1086/673168}

\bibitem[{D.~A. {Caldwell} {et~al.}(2020){Caldwell}, {Tenenbaum}, {Twicken}, {Jenkins}, {Ting}, {Smith}, {Hedges}, {Fausnaugh}, {Rose}, \& {Burke}}]{Caldwell2020}
{Caldwell}, D.~A., {Tenenbaum}, P., {Twicken}, J.~D., {et~al.} 2020, \bibinfo{title}{{TESS Science Processing Operations Center FFI Target List Products},} Research Notes of the American Astronomical Society, 4, 201, \dodoi{10.3847/2515-5172/abc9b3}

\bibitem[{O. {Cohen} {et~al.}(2010){Cohen}, {Drake}, {Kashyap}, {Sokolov}, \& {Gombosi}}]{Cohen2010}
{Cohen}, O., {Drake}, J.~J., {Kashyap}, V.~L., {Sokolov}, I.~V., \& {Gombosi}, T.~I. 2010, \bibinfo{title}{{The Impact of Hot Jupiters on the Spin-down of their Host Stars},} \apjl, 723, L64, \dodoi{10.1088/2041-8205/723/1/L64}

\bibitem[{A. {Collier Cameron} \& M. {Jardine}(2018){Collier Cameron} \& {Jardine}}]{Cameron2018}
{Collier Cameron}, A., \& {Jardine}, M. 2018, \bibinfo{title}{{Hierarchical Bayesian calibration of tidal orbit decay rates among hot Jupiters},} \mnras, 476, 2542, \dodoi{10.1093/mnras/sty292}

\bibitem[{A. {Collier Cameron} \& J. {Li}(1994){Collier Cameron} \& {Li}}]{Cameron1994}
{Collier Cameron}, A., \& {Li}, J. 1994, \bibinfo{title}{{Magnetic braking of G and K dwarfs without core-envelope decoupling.},} \mnras, 269, 1099, \dodoi{10.1093/mnras/269.4.1099}

\bibitem[{D. {Cont} {et~al.}(2025){Cont}, {Nortmann}, {Lesjak}, {Yan}, {Shulyak}, {Lavail}, {Stangret}, {Pall{\'e}}, {Amado}, {Caballero}, {Hatzes}, {Henning}, {Piskunov}, {Quirrenbach}, {Reiners}, {Ribas}, {Ag{\"u}{\'\i} Fern{\'a}ndez}, {Ak{\i}n}, {Boldt-Christmas}, {Chaturvedi}, {Czesla}, {Hahlin}, {Heng}, {Kochukhov}, {Marquart}, {Molaverdikhani}, {Montes}, {Morello}, {Nagel}, {Orell-Miquel}, {Rains}, {Rengel}, {Schweitzer}, {S{\'a}nchez-L{\'o}pez}, \& {Seemann}}]{Cont2025}
{Cont}, D., {Nortmann}, L., {Lesjak}, F., {et~al.} 2025, \bibinfo{title}{{Retrieving day- and nightside atmospheric properties of the ultra-hot Jupiter TOI-2109b. Detection of Fe and CO emission lines and evidence for inefficient heat transport},} arXiv e-prints, arXiv:2504.15757.
\newblock \doarXiv{2504.15757}

\bibitem[{G.~H. {Darwin}(1879){Darwin}}]{Darwin1879}
{Darwin}, G.~H. 1879, \bibinfo{title}{{On the Bodily Tides of Viscous and Semi-Elastic Spheroids, and on the Ocean Tides upon a Yielding Nucleus},} Philosophical Transactions of the Royal Society of London Series I, 170, 1

\bibitem[{R.~I. {Dawson} \& J.~A. {Johnson}(2018){Dawson} \& {Johnson}}]{Dawson2018}
{Dawson}, R.~I., \& {Johnson}, J.~A. 2018, \bibinfo{title}{{Origins of Hot Jupiters},} \araa, 56, 175, \dodoi{10.1146/annurev-astro-081817-051853}

\bibitem[{K. De {et~al.}(2023)De, MacLeod, Karambelkar, Jencson, Chakrabarty, Conroy, Dekany, Eilers, Graham, Hillenbrand, Kara, Kasliwal, Kulkarni, Lau, Loeb, Masci, Medford, Meisner, Patel, \& Vanderburg}]{De2023}
De, K., MacLeod, M., Karambelkar, V., {et~al.} 2023, \bibinfo{title}{An infrared transient from a star engulfing a planet,} Nature, 617, 55, \dodoi{10.1038/s41586-023-05842-x}

\bibitem[{L. {Delrez} {et~al.}(2016){Delrez}, {Santerne}, {Almenara}, {Anderson}, {Collier-Cameron}, {D{\'\i}az}, {Gillon}, {Hellier}, {Jehin}, {Lendl}, {Maxted}, {Neveu-VanMalle}, {Pepe}, {Pollacco}, {Queloz}, {S{\'e}gransan}, {Smalley}, {Smith}, {Triaud}, {Udry}, {Van Grootel}, \& {West}}]{Delrez2016}
{Delrez}, L., {Santerne}, A., {Almenara}, J.~M., {et~al.} 2016, \bibinfo{title}{{WASP-121 b: a hot Jupiter close to tidal disruption transiting an active F star},} \mnras, 458, 4025, \dodoi{10.1093/mnras/stw522}

\bibitem[{C.~D. {Duguid} {et~al.}(2024){Duguid}, {de Vries}, {Lecoanet}, \& {Barker}}]{Duguid2024}
{Duguid}, C.~D., {de Vries}, N.~B., {Lecoanet}, D., \& {Barker}, A.~J. 2024, \bibinfo{title}{{An Efficient Tidal Dissipation Mechanism via Stellar Magnetic Fields},} \apjl, 966, L14, \dodoi{10.3847/2041-8213/ad3c40}

\bibitem[{M. {Efroimsky}(2012){Efroimsky}}]{Efroimsky2012}
{Efroimsky}, M. 2012, \bibinfo{title}{{Tidal Dissipation Compared to Seismic Dissipation: In Small Bodies, Earths, and Super-Earths},} \apj, 746, 150, \dodoi{10.1088/0004-637X/746/2/150}

\bibitem[{R. {Essick} \& N.~N. {Weinberg}(2016){Essick} \& {Weinberg}}]{Essick2016}
{Essick}, R., \& {Weinberg}, N.~N. 2016, \bibinfo{title}{{Orbital Decay of Hot Jupiters Due to Nonlinear Tidal Dissipation within Solar-type Hosts},} \apj, 816, 18, \dodoi{10.3847/0004-637X/816/1/18}

\bibitem[{L. {Fellay} {et~al.}(2023){Fellay}, {Pezzotti}, {Buldgen}, {Eggenberger}, \& {Bolmont}}]{Fellay2023}
{Fellay}, L., {Pezzotti}, C., {Buldgen}, G., {Eggenberger}, P., \& {Bolmont}, E. 2023, \bibinfo{title}{{Constraints on planetary tidal dissipation from a detailed study of Kepler 91b},} \aap, 669, A2, \dodoi{10.1051/0004-6361/202243621}

\bibitem[{A.~J. {Finley} \& S.~P. {Matt}(2017){Finley} \& {Matt}}]{Finley2017}
{Finley}, A.~J., \& {Matt}, S.~P. 2017, \bibinfo{title}{{The Effect of Combined Magnetic Geometries on Thermally Driven Winds. I. Interaction of Dipolar and Quadrupolar Fields},} \apj, 845, 46, \dodoi{10.3847/1538-4357/aa7fb9}

\bibitem[{F. {Gallet} {et~al.}(2017){Gallet}, {Bolmont}, {Mathis}, {Charbonnel}, \& {Amard}}]{Gallet2017}
{Gallet}, F., {Bolmont}, E., {Mathis}, S., {Charbonnel}, C., \& {Amard}, L. 2017, \bibinfo{title}{{Tidal dissipation in rotating low-mass stars and implications for the orbital evolution of close-in planets. I. From the PMS to the RGB at solar metallicity},} \aap, 604, A112, \dodoi{10.1051/0004-6361/201730661}

\bibitem[{B.~S. {Gaudi} {et~al.}(2017){Gaudi}, {Stassun}, {Collins}, {Beatty}, {Zhou}, {Latham}, {Bieryla}, {Eastman}, {Siverd}, {Crepp}, {Gonzales}, {Stevens}, {Buchhave}, {Pepper}, {Johnson}, {Colon}, {Jensen}, {Rodriguez}, {Bozza}, {Novati}, {D'Ago}, {Dumont}, {Ellis}, {Gaillard}, {Jang-Condell}, {Kasper}, {Fukui}, {Gregorio}, {Ito}, {Kielkopf}, {Manner}, {Matt}, {Narita}, {Oberst}, {Reed}, {Scarpetta}, {Stephens}, {Yeigh}, {Zambelli}, {Fulton}, {Howard}, {James}, {Penny}, {Bayliss}, {Curtis}, {Depoy}, {Esquerdo}, {Gould}, {Joner}, {Kuhn}, {Labadie-Bartz}, {Lund}, {Marshall}, {McLeod}, {Pogge}, {Relles}, {Stockdale}, {Tan}, {Trueblood}, \& {Trueblood}}]{Gaudi2017}
{Gaudi}, B.~S., {Stassun}, K.~G., {Collins}, K.~A., {et~al.} 2017, \bibinfo{title}{{A giant planet undergoing extreme-ultraviolet irradiation by its hot massive-star host},} \nat, 546, 514, \dodoi{10.1038/nature22392}

\bibitem[{P. {Goldreich} \& S. {Soter}(1966){Goldreich} \& {Soter}}]{Goldreich1966}
{Goldreich}, P., \& {Soter}, S. 1966, \bibinfo{title}{{Q in the Solar System},} icarus, 5, 375, \dodoi{10.1016/0019-1035(66)90051-0}

\bibitem[{A.~V. {Goyal} \& S. {Wang}(2025){Goyal} \& {Wang}}]{Goyal2025}
{Goyal}, A.~V., \& {Wang}, S. 2025, \bibinfo{title}{{Statistical Reevaluation of the Ultra-short-period Planet Classification Boundary: Smaller Planets within 1 Day, Larger Period Ratios below 2 Days},} \aj, 169, 191, \dodoi{10.3847/1538-3881/adb487}

\bibitem[{P.-G. {Gu} {et~al.}(2003){Gu}, {Lin}, \& {Bodenheimer}}]{Gu2003}
{Gu}, P.-G., {Lin}, D. N.~C., \& {Bodenheimer}, P.~H. 2003, \bibinfo{title}{{The Effect of Tidal Inflation Instability on the Mass and Dynamical Evolution of Extrasolar Planets with Ultrashort Periods},} \apj, 588, 509, \dodoi{10.1086/373920}

\bibitem[{M. {Guenel} {et~al.}(2014){Guenel}, {Mathis}, \& {Remus}}]{Guenel2014}
{Guenel}, M., {Mathis}, S., \& {Remus}, F. 2014, \bibinfo{title}{{Unravelling tidal dissipation in gaseous giant planets},} \aap, 566, L9, \dodoi{10.1051/0004-6361/201424010}

\bibitem[{S.~R. {Hagey} {et~al.}(2022){Hagey}, {Edwards}, \& {Boley}}]{Hagey2022}
{Hagey}, S.~R., {Edwards}, B., \& {Boley}, A.~C. 2022, \bibinfo{title}{{Evidence of Long-term Period Variations in the Exoplanet Transit Database (ETD)},} \aj, 164, 220, \dodoi{10.3847/1538-3881/ac959a}

\bibitem[{J.~H. {Hamer} \& K.~C. {Schlaufman}(2020){Hamer} \& {Schlaufman}}]{Hamer2020}
{Hamer}, J.~H., \& {Schlaufman}, K.~C. 2020, \bibinfo{title}{{Ultra-short-period Planets Are Stable against Tidal Inspiral},} \aj, 160, 138, \dodoi{10.3847/1538-3881/aba74f}

\bibitem[{J.-V. {Harre} \& A.~M.~S. {Smith}(2023){Harre} \& {Smith}}]{Harre2023}
{Harre}, J.-V., \& {Smith}, A. M.~S. 2023, \bibinfo{title}{{The Apparent Tidal Decay of WASP-4 b Can Be Explained by the R{\o}mer Effect},} Universe, 9, 506, \dodoi{10.3390/universe9120506}

\bibitem[{J.~V. {Harre} {et~al.}(2024){Harre}, {Smith}, {Barros}, {Singh}, {Korth}, {Brandeker}, {Collier Cameron}, {Lendl}, {Wilson}, {Borsato}, {Csizmadia}, {Cabrera}, {Parviainen}, {Correia}, {Akinsanmi}, {Rosario}, {Leonardi}, {Serrano}, {Alibert}, {Alonso}, {Asquier}, {B{\'a}rczy}, {Barrado Navascues}, {Baumjohann}, {Benz}, {Billot}, {Broeg}, {Busch}, {Cubillos}, {Davies}, {Deleuil}, {Deline}, {Delrez}, {Demangeon}, {Demory}, {Derekas}, {Edwards}, {Ehrenreich}, {Erikson}, {Fortier}, {Fossati}, {Fridlund}, {Gandolfi}, {Gazeas}, {Gillon}, {G{\"u}del}, {G{\"u}nther}, {Heitzmann}, {Helling}, {Isaak}, {Kiss}, {Lam}, {Laskar}, {Lecavelier des Etangs}, {Magrin}, {Maxted}, {Mer{\'\i}n}, {Mordasini}, {Nascimbeni}, {Olofsson}, {Ottensamer}, {Pagano}, {Pall{\'e}}, {Peter}, {Piazza}, {Piotto}, {Pollacco}, {Queloz}, {Ragazzoni}, {Rando}, {Rauer}, {Ribas}, {Santos}, {Scandariato}, {S{\'e}gransan}, {Simon}, {Sousa}, {Stalport}, {Sulis}, {Szab{\'o}}, {Udry}, {Ulmer}, {Van Grootel}, {Venturini}, {Villaver}, {Viotto},
  {Walton}, {West}, \& {Westerdorff}}]{Harre2024}
{Harre}, J.~V., {Smith}, A.~M.~S., {Barros}, S.~C.~C., {et~al.} 2024, \bibinfo{title}{{Hints of a close outer companion to the ultra-hot Jupiter TOI-2109 b},} \aap, 692, A254, \dodoi{10.1051/0004-6361/202451068}

\bibitem[{Q. {Hou} \& X. {Wei}(2022){Hou} \& {Wei}}]{Hou2022}
{Hou}, Q., \& {Wei}, X. 2022, \bibinfo{title}{{Why hot Jupiters can be large but not too large},} \mnras, 511, 3133, \dodoi{10.1093/mnras/stac169}

\bibitem[{A.~W. {Howard} {et~al.}(2012){Howard}, {Marcy}, {Bryson}, {Jenkins}, {Rowe}, {Batalha}, {Borucki}, {Koch}, {Dunham}, {Gautier}, {Van Cleve}, {Cochran}, {Latham}, {Lissauer}, {Torres}, {Brown}, {Gilliland}, {Buchhave}, {Caldwell}, {Christensen-Dalsgaard}, {Ciardi}, {Fressin}, {Haas}, {Howell}, {Kjeldsen}, {Seager}, {Rogers}, {Sasselov}, {Steffen}, {Basri}, {Charbonneau}, {Christiansen}, {Clarke}, {Dupree}, {Fabrycky}, {Fischer}, {Ford}, {Fortney}, {Tarter}, {Girouard}, {Holman}, {Johnson}, {Klaus}, {Machalek}, {Moorhead}, {Morehead}, {Ragozzine}, {Tenenbaum}, {Twicken}, {Quinn}, {Isaacson}, {Shporer}, {Lucas}, {Walkowicz}, {Welsh}, {Boss}, {Devore}, {Gould}, {Smith}, {Morris}, {Prsa}, {Morton}, {Still}, {Thompson}, {Mullally}, {Endl}, \& {MacQueen}}]{Howard2012}
{Howard}, A.~W., {Marcy}, G.~W., {Bryson}, S.~T., {et~al.} 2012, \bibinfo{title}{{Planet Occurrence within 0.25 AU of Solar-type Stars from Kepler},} \apjs, 201, 15, \dodoi{10.1088/0067-0049/201/2/15}

\bibitem[{S. {Hoyer} {et~al.}(2020){Hoyer}, {Guterman}, {Demangeon}, {Sousa}, {Deleuil}, {Meunier}, \& {Benz}}]{Hoyer2020}
{Hoyer}, S., {Guterman}, P., {Demangeon}, O., {et~al.} 2020, \bibinfo{title}{{Expected performances of the Characterising Exoplanet Satellite (CHEOPS). III. Data reduction pipeline: architecture and simulated performances},} \aap, 635, A24, \dodoi{10.1051/0004-6361/201936325}

\bibitem[{P. {Hut}(1981){Hut}}]{Hut1981}
{Hut}, P. 1981, \bibinfo{title}{{Tidal evolution in close binary systems.},} \aap, 99, 126

\bibitem[{P.~B. {Ivanov} {et~al.}(2013){Ivanov}, {Papaloizou}, \& {Chernov}}]{Ivanov2013}
{Ivanov}, P.~B., {Papaloizou}, J.~C.~B., \& {Chernov}, S.~V. 2013, \bibinfo{title}{{A unified normal mode approach to dynamic tides and its application to rotating Sun-like stars},} \mnras, 432, 2339, \dodoi{10.1093/mnras/stt595}

\bibitem[{E.~S. {Ivshina} \& J.~N. {Winn}(2022){Ivshina} \& {Winn}}]{Ivshina2022}
{Ivshina}, E.~S., \& {Winn}, J.~N. 2022, \bibinfo{title}{{TESS Transit Timing of Hundreds of Hot Jupiters},} \apjs, 259, 62, \dodoi{10.3847/1538-4365/ac545b}

\bibitem[{B. {Jackson} {et~al.}(2016){Jackson}, {Jensen}, {Peacock}, {Arras}, \& {Penev}}]{Jackson2016}
{Jackson}, B., {Jensen}, E., {Peacock}, S., {Arras}, P., \& {Penev}, K. 2016, \bibinfo{title}{{Tidal decay and stable Roche-lobe overflow of short-period gaseous exoplanets},} Celestial Mechanics and Dynamical Astronomy, 126, 227, \dodoi{10.1007/s10569-016-9704-1}

\bibitem[{C.~P. {Johnstone} {et~al.}(2015){Johnstone}, {G{\"u}del}, {L{\"u}ftinger}, {Toth}, \& {Brott}}]{Johnstone2015}
{Johnstone}, C.~P., {G{\"u}del}, M., {L{\"u}ftinger}, T., {Toth}, G., \& {Brott}, I. 2015, \bibinfo{title}{{Stellar winds on the main-sequence. I. Wind model},} \aap, 577, A27, \dodoi{10.1051/0004-6361/201425300}

\bibitem[{K. {Jones} {et~al.}(2022){Jones}, {Morris}, {Demory}, {Heng}, {Hooton}, {Billot}, {Ehrenreich}, {Hoyer}, {Simon}, {Lendl}, {Demangeon}, {Sousa}, {Bonfanti}, {Wilson}, {Salmon}, {Csizmadia}, {Parviainen}, {Bruno}, {Alibert}, {Alonso}, {Anglada}, {B{\'a}rczy}, {Barrado}, {Barros}, {Baumjohann}, {Beck}, {Beck}, {Benz}, {Bonfils}, {Brandeker}, {Broeg}, {Cabrera}, {Charnoz}, {Collier Cameron}, {Davies}, {Deleuil}, {Deline}, {Delrez}, {Erikson}, {Fortier}, {Fossati}, {Fridlund}, {Gandolfi}, {Gillon}, {G{\"u}del}, {Isaak}, {Kiss}, {Laskar}, {Lecavelier des Etangs}, {Lovis}, {Magrin}, {Maxted}, {Nascimbeni}, {Olofsson}, {Ottensamer}, {Pagano}, {Pall{\'e}}, {Peter}, {Piotto}, {Pollacco}, {Queloz}, {Ragazzoni}, {Rando}, {Ratti}, {Rauer}, {Reimers}, {Ribas}, {Santos}, {Scandariato}, {S{\'e}gransan}, {Smith}, {Steller}, {Szab{\'o}}, {Thomas}, {Udry}, {Van Grootel}, {Walter}, {Walton}, \& {Wang Jungo}}]{Jones2022}
{Jones}, K., {Morris}, B.~M., {Demory}, B.~O., {et~al.} 2022, \bibinfo{title}{{The stable climate of KELT-9b},} \aap, 666, A118, \dodoi{10.1051/0004-6361/202243823}

\bibitem[{Z. Kopal(1960)Kopal}]{Kopal1960}
Kopal, Z. 1960, Figures of Equilibrium of Celestial Bodies: With Emphasis on Problems of Motion of Artificial Satellites (Madison, WI: University of Wisconsin Press), 135

\bibitem[{K. {Kotorashvili} \& E.~G. {Blackman}(2024){Kotorashvili} \& {Blackman}}]{Kotorashvili2024}
{Kotorashvili}, K., \& {Blackman}, E.~G. 2024, \bibinfo{title}{{Tidally Delayed Spin-Down of Very Low Mass Stars},} arXiv e-prints, arXiv:2411.17916.
\newblock \doarXiv{2411.17916}

\bibitem[{V. {Lainey} {et~al.}(2012){Lainey}, {Karatekin}, {Desmars}, {Charnoz}, {Arlot}, {Emelyanov}, {Le Poncin-Lafitte}, {Mathis}, {Remus}, {Tobie}, \& {Zahn}}]{Lainey2012}
{Lainey}, V., {Karatekin}, {\"O}., {Desmars}, J., {et~al.} 2012, \bibinfo{title}{{Strong Tidal Dissipation in Saturn and Constraints on Enceladus' Thermal State from Astrometry},} \apj, 752, 14, \dodoi{10.1088/0004-637X/752/1/14}

\bibitem[{A.~F. {Lanza}(2010){Lanza}}]{Lanza2010}
{Lanza}, A.~F. 2010, \bibinfo{title}{{Hot Jupiters and the evolution of stellar angular momentum},} \aap, 512, A77, \dodoi{10.1051/0004-6361/200912789}

\bibitem[{A.~F. {Lanza} {et~al.}(2011){Lanza}, {Damiani}, \& {Gandolfi}}]{Lanza2011}
{Lanza}, A.~F., {Damiani}, C., \& {Gandolfi}, D. 2011, \bibinfo{title}{{Constraining tidal dissipation in F-type main-sequence stars: the case of CoRoT-11},} \aap, 529, A50, \dodoi{10.1051/0004-6361/201016144}

\bibitem[{Y.~A. {Lazovik}(2021){Lazovik}}]{Lazovik2021}
{Lazovik}, Y.~A. 2021, \bibinfo{title}{{Tidal migration of hot Jupiters: introducing the impact of gravity wave dissipation},} \mnras, 508, 3408, \dodoi{10.1093/mnras/stab2768}

\bibitem[{J. {Leconte} {et~al.}(2010){Leconte}, {Chabrier}, {Baraffe}, \& {Levrard}}]{Leconte2010}
{Leconte}, J., {Chabrier}, G., {Baraffe}, I., \& {Levrard}, B. 2010, \bibinfo{title}{{Is tidal heating sufficient to explain bloated exoplanets? Consistent calculations accounting for finite initial eccentricity},} \aap, 516, A64, \dodoi{10.1051/0004-6361/201014337}

\bibitem[{P. {Leonardi} {et~al.}(2024){Leonardi}, {Nascimbeni}, {Granata}, {Malavolta}, {Borsato}, {Biazzo}, {Lanza}, {Desidera}, {Piotto}, {Nardiello}, {Damasso}, {Cunial}, \& {Bedin}}]{Leonardi2024}
{Leonardi}, P., {Nascimbeni}, V., {Granata}, V., {et~al.} 2024, \bibinfo{title}{{TASTE. V. A new ground-based investigation of orbital decay in the ultra-hot Jupiter WASP-12b},} \aap, 686, A84, \dodoi{10.1051/0004-6361/202348363}

\bibitem[{B. {Levrard} {et~al.}(2009){Levrard}, {Winisdoerffer}, \& {Chabrier}}]{Levrard2009}
{Levrard}, B., {Winisdoerffer}, C., \& {Chabrier}, G. 2009, \bibinfo{title}{{Falling Transiting Extrasolar Giant Planets},} \apjl, 692, L9, \dodoi{10.1088/0004-637X/692/1/L9}

\bibitem[{ {Lightkurve Collaboration} {et~al.}(2018){Lightkurve Collaboration}, {Cardoso}, {Hedges}, {Gully-Santiago}, {Saunders}, {Cody}, {Barclay}, {Hall}, {Sagear}, {Turtelboom}, {Zhang}, {Tzanidakis}, {Mighell}, {Coughlin}, {Bell}, {Berta-Thompson}, {Williams}, {Dotson}, \& {Barentsen}}]{Lightkurve2018}
{Lightkurve Collaboration}, {Cardoso}, J.~V.~d.~M., {Hedges}, C., {et~al.} 2018, \bibinfo{title}{{Lightkurve: Kepler and TESS time series analysis in Python},}, Astrophysics Source Code Library \doeprint{1812.013}

\bibitem[{Y. {Lin} \& G.~I. {Ogilvie}(2018){Lin} \& {Ogilvie}}]{Lin2018}
{Lin}, Y., \& {Ogilvie}, G.~I. 2018, \bibinfo{title}{{Tidal dissipation in rotating fluid bodies: the presence of a magnetic field},} \mnras, 474, 1644, \dodoi{10.1093/mnras/stx2764}

\bibitem[{A.~E.~H. {Love}(1927){Love}}]{Love1927}
{Love}, A.~E.~H. 1927, {A Treatise on the Mathematical Theory of Elasticity}, Vol.~4 (Cambridge University Press)

\bibitem[{L. {Ma} \& J. {Fuller}(2021){Ma} \& {Fuller}}]{Ma2021}
{Ma}, L., \& {Fuller}, J. 2021, \bibinfo{title}{{Orbital Decay of Short-period Exoplanets via Tidal Resonance Locking},} \apj, 918, 16, \dodoi{10.3847/1538-4357/ac088e}

\bibitem[{G. {Maciejewski} {et~al.}(2021){Maciejewski}, {Fern{\'a}ndez}, {Aceituno}, {Ramos}, {Dimitrov}, {Donchev}, \& {Ohlert}}]{Maciejewski2021}
{Maciejewski}, G., {Fern{\'a}ndez}, M., {Aceituno}, F., {et~al.} 2021, \bibinfo{title}{{Revisiting TrES-5 b: departure from a linear ephemeris instead of short-period transit timing variation},} \aap, 656, A88, \dodoi{10.1051/0004-6361/202142424}

\bibitem[{G. {Maciejewski} {et~al.}(2016){Maciejewski}, {Dimitrov}, {Fern{\'a}ndez}, {Sota}, {Nowak}, {Ohlert}, {Nikolov}, {Bukowiecki}, {Hinse}, {Pall{\'e}}, {Tingley}, {Kjurkchieva}, {Lee}, \& {Lee}}]{Maciejewsky2016}
{Maciejewski}, G., {Dimitrov}, D., {Fern{\'a}ndez}, M., {et~al.} 2016, \bibinfo{title}{{Departure from the constant-period ephemeris for the transiting exoplanet WASP-12},} \aap, 588, L6, \dodoi{10.1051/0004-6361/201628312}

\bibitem[{K. {Mandel} \& E. {Agol}(2002){Mandel} \& {Agol}}]{Mandel2002}
{Mandel}, K., \& {Agol}, E. 2002, \bibinfo{title}{{Analytic Light Curves for Planetary Transit Searches},} \apjl, 580, L171, \dodoi{10.1086/345520}

\bibitem[{S. {Mathis}(2015){Mathis}}]{Mathis2015}
{Mathis}, S. 2015, \bibinfo{title}{{Variation of tidal dissipation in the convective envelope of low-mass stars along their evolution},} \aap, 580, L3, \dodoi{10.1051/0004-6361/201526472}

\bibitem[{M. {Mayor} \& D. {Queloz}(1995){Mayor} \& {Queloz}}]{Mayor1995}
{Mayor}, M., \& {Queloz}, D. 1995, \bibinfo{title}{{A Jupiter-mass companion to a solar-type star},} \nat, 378, 355, \dodoi{10.1038/378355a0}

\bibitem[{S. {Millholland}(2019){Millholland}}]{Millholland2019}
{Millholland}, S. 2019, \bibinfo{title}{{Tidally Induced Radius Inflation of Sub-Neptunes},} \apj, 886, 72, \dodoi{10.3847/1538-4357/ab4c3f}

\bibitem[{M. {Mol Lous} \& Y. {Miguel}(2020){Mol Lous} \& {Miguel}}]{MolLous2020}
{Mol Lous}, M., \& {Miguel}, Y. 2020, \bibinfo{title}{{Inflation of migrated hot Jupiters},} \mnras, 495, 2994, \dodoi{10.1093/mnras/staa1405}

\bibitem[{S. {M{\"u}ller} {et~al.}(2020){M{\"u}ller}, {Helled}, \& {Cumming}}]{Muller2020}
{M{\"u}ller}, S., {Helled}, R., \& {Cumming}, A. 2020, \bibinfo{title}{{The challenge of forming a fuzzy core in Jupiter},} \aap, 638, A121, \dodoi{10.1051/0004-6361/201937376}

\bibitem[{C.~D. {Murray} \& S.~F. {Dermott}(2000){Murray} \& {Dermott}}]{Murray2000}
{Murray}, C.~D., \& {Dermott}, S.~F. 2000, {Solar System Dynamics (New York: Cambridge Univ. Press)}

\bibitem[{ {NASA Exoplanet Archive}(2025){NASA Exoplanet Archive}}]{NASAExoArch}
{NASA Exoplanet Archive}. 2025, \bibinfo{title}{Planetary Systems,}, Version: 2025-04-27 21:00 NExScI-Caltech/IPAC, \dodoi{10.26133/NEA12}

\bibitem[{T.~E. {Oberst} {et~al.}(2017){Oberst}, {Rodriguez}, {Col{\'o}n}, {Angerhausen}, {Bieryla}, {Ngo}, {Stevens}, {Stassun}, {Gaudi}, {Pepper}, {Penev}, {Mawet}, {Latham}, {Heintz}, {Osei}, {Collins}, {Kielkopf}, {Visgaitis}, {Reed}, {Escamilla}, {Yazdi}, {McLeod}, {Lunsford}, {Spencer}, {Joner}, {Gregorio}, {Gaillard}, {Matt}, {Dumont}, {Stephens}, {Cohen}, {Jensen}, {Calchi Novati}, {Bozza}, {Labadie-Bartz}, {Siverd}, {Lund}, {Beatty}, {Eastman}, {Penny}, {Manner}, {Zambelli}, {Fulton}, {Stockdale}, {DePoy}, {Marshall}, {Pogge}, {Gould}, {Trueblood}, \& {Trueblood}}]{Oberst2017}
{Oberst}, T.~E., {Rodriguez}, J.~E., {Col{\'o}n}, K.~D., {et~al.} 2017, \bibinfo{title}{{KELT-16b: A Highly Irradiated, Ultra-short Period Hot Jupiter Nearing Tidal Disruption},} \aj, 153, 97, \dodoi{10.3847/1538-3881/153/3/97}

\bibitem[{G.~I. {Ogilvie}(2013){Ogilvie}}]{Ogilvie2013}
{Ogilvie}, G.~I. 2013, \bibinfo{title}{{Tides in rotating barotropic fluid bodies: the contribution of inertial waves and the role of internal structure},} \mnras, 429, 613, \dodoi{10.1093/mnras/sts362}

\bibitem[{G.~I. {Ogilvie} \& D.~N.~C. {Lin}(2007){Ogilvie} \& {Lin}}]{Ogilvie2007}
{Ogilvie}, G.~I., \& {Lin}, D.~N.~C. 2007, \bibinfo{title}{{Tidal Dissipation in Rotating Solar-Type Stars},} \apj, 661, 1180, \dodoi{10.1086/515435}

\bibitem[{S. {Oh} {et~al.}(2018){Oh}, {Price-Whelan}, {Brewer}, {Hogg}, {Spergel}, \& {Myles}}]{Oh2018}
{Oh}, S., {Price-Whelan}, A.~M., {Brewer}, J.~M., {et~al.} 2018, \bibinfo{title}{{Kronos and Krios: Evidence for Accretion of a Massive, Rocky Planetary System in a Comoving Pair of Solar-type Stars},} \apj, 854, 138, \dodoi{10.3847/1538-4357/aaab4d}

\bibitem[{V. {Parmentier} {et~al.}(2018){Parmentier}, {Line}, {Bean}, {Mansfield}, {Kreidberg}, {Lupu}, {Visscher}, {D{\'e}sert}, {Fortney}, {Deleuil}, {Arcangeli}, {Showman}, \& {Marley}}]{Parmentier2018}
{Parmentier}, V., {Line}, M.~R., {Bean}, J.~L., {et~al.} 2018, \bibinfo{title}{{From thermal dissociation to condensation in the atmospheres of ultra hot Jupiters: WASP-121b in context},} \aap, 617, A110, \dodoi{10.1051/0004-6361/201833059}

\bibitem[{H. {Parviainen}(2015){Parviainen}}]{Parviainen2015}
{Parviainen}, H. 2015, \bibinfo{title}{{PYTRANSIT: fast and easy exoplanet transit modelling in PYTHON},} \mnras, 450, 3233, \dodoi{10.1093/mnras/stv894}

\bibitem[{K.~C. {Patra} {et~al.}(2017){Patra}, {Winn}, {Holman}, {Yu}, {Deming}, \& {Dai}}]{Patra2017}
{Patra}, K.~C., {Winn}, J.~N., {Holman}, M.~J., {et~al.} 2017, \bibinfo{title}{{The Apparently Decaying Orbit of WASP-12b},} \aj, 154, 4, \dodoi{10.3847/1538-3881/aa6d75}

\bibitem[{K.~C. {Patra} {et~al.}(2020){Patra}, {Winn}, {Holman}, {Gillon}, {Burdanov}, {Jehin}, {Delrez}, {Pozuelos}, {Barkaoui}, {Benkhaldoun}, {Narita}, {Fukui}, {Kusakabe}, {Kawauchi}, {Terada}, {Bouma}, {Weinberg}, \& {Broome}}]{Patra2020}
{Patra}, K.~C., {Winn}, J.~N., {Holman}, M.~J., {et~al.} 2020, \bibinfo{title}{{The Continuing Search for Evidence of Tidal Orbital Decay of Hot Jupiters},} \aj, 159, 150, \dodoi{10.3847/1538-3881/ab7374}

\bibitem[{K. {Penev} {et~al.}(2018){Penev}, {Bouma}, {Winn}, \& {Hartman}}]{Penev2018}
{Penev}, K., {Bouma}, L.~G., {Winn}, J.~N., \& {Hartman}, J.~D. 2018, \bibinfo{title}{{Empirical Tidal Dissipation in Exoplanet Hosts From Tidal Spin-up},} \aj, 155, 165, \dodoi{10.3847/1538-3881/aaaf71}

\bibitem[{K. {Penev} {et~al.}(2016){Penev}, {Hartman}, {Bakos}, {Ciceri}, {Brahm}, {Bayliss}, {Bento}, {Jord{\'a}n}, {Csubry}, {Bhatti}, {de Val-Borro}, {Espinoza}, {Zhou}, {Mancini}, {Rabus}, {Suc}, {Henning}, {Schmidt}, {Noyes}, {L{\'a}z{\'a}r}, {Papp}, \& {S{\'a}ri}}]{Penev2016}
{Penev}, K., {Hartman}, J.~D., {Bakos}, G.~{\'A}., {et~al.} 2016, \bibinfo{title}{{HATS-18b: An Extreme Short-period Massive Transiting Planet Spinning Up Its Star},} \aj, 152, 127, \dodoi{10.3847/0004-6256/152/5/127}

\bibitem[{R. {Petrucci} {et~al.}(2020){Petrucci}, {Jofr{\'e}}, {G{\'o}mez Maqueo Chew}, {Hinse}, {Ma{\v{s}}ek}, {Tan}, \& {G{\'o}mez}}]{Petrucci2020}
{Petrucci}, R., {Jofr{\'e}}, E., {G{\'o}mez Maqueo Chew}, Y., {et~al.} 2020, \bibinfo{title}{{Discarding orbital decay in WASP-19b after one decade of transit observations},} \mnras, 491, 1243, \dodoi{10.1093/mnras/stz3034}

\bibitem[{D.~L. {Pollacco} {et~al.}(2006){Pollacco}, {Skillen}, {Collier Cameron}, {Christian}, {Hellier}, {Irwin}, {Lister}, {Street}, {West}, {Anderson}, {Clarkson}, {Deeg}, {Enoch}, {Evans}, {Fitzsimmons}, {Haswell}, {Hodgkin}, {Horne}, {Kane}, {Keenan}, {Maxted}, {Norton}, {Osborne}, {Parley}, {Ryans}, {Smalley}, {Wheatley}, \& {Wilson}}]{Pollacco2006}
{Pollacco}, D.~L., {Skillen}, I., {Collier Cameron}, A., {et~al.} 2006, \bibinfo{title}{{The WASP Project and the SuperWASP Cameras},} \pasp, 118, 1407, \dodoi{10.1086/508556}

\bibitem[{H. {Rein} \& S.~F. {Liu}(2012){Rein} \& {Liu}}]{rebound}
{Rein}, H., \& {Liu}, S.~F. 2012, \bibinfo{title}{{REBOUND: an open-source multi-purpose N-body code for collisional dynamics},} \aap, 537, A128, \dodoi{10.1051/0004-6361/201118085}

\bibitem[{H. {Rein} \& D. {Tamayo}(2015){Rein} \& {Tamayo}}]{reboundwhfast}
{Rein}, H., \& {Tamayo}, D. 2015, \bibinfo{title}{{WHFAST: a fast and unbiased implementation of a symplectic Wisdom-Holman integrator for long-term gravitational simulations},} \mnras, 452, 376, \dodoi{10.1093/mnras/stv1257}

\bibitem[{G.~R. {Ricker} {et~al.}(2015){Ricker}, {Winn}, {Vanderspek}, {Latham}, {Bakos}, {Bean}, {Berta-Thompson}, {Brown}, {Buchhave}, {Butler}, {Butler}, {Chaplin}, {Charbonneau}, {Christensen-Dalsgaard}, {Clampin}, {Deming}, {Doty}, {De Lee}, {Dressing}, {Dunham}, {Endl}, {Fressin}, {Ge}, {Henning}, {Holman}, {Howard}, {Ida}, {Jenkins}, {Jernigan}, {Johnson}, {Kaltenegger}, {Kawai}, {Kjeldsen}, {Laughlin}, {Levine}, {Lin}, {Lissauer}, {MacQueen}, {Marcy}, {McCullough}, {Morton}, {Narita}, {Paegert}, {Palle}, {Pepe}, {Pepper}, {Quirrenbach}, {Rinehart}, {Sasselov}, {Sato}, {Seager}, {Sozzetti}, {Stassun}, {Sullivan}, {Szentgyorgyi}, {Torres}, {Udry}, \& {Villasenor}}]{Ricker2015}
{Ricker}, G.~R., {Winn}, J.~N., {Vanderspek}, R., {et~al.} 2015, \bibinfo{title}{{Transiting Exoplanet Survey Satellite (TESS)},} JATIS, 1, 014003, \dodoi{10.1117/1.JATIS.1.1.014003}

\bibitem[{M. {Rieutord} \& L. {Valdettaro}(2010){Rieutord} \& {Valdettaro}}]{Rieutord2010}
{Rieutord}, M., \& {Valdettaro}, L. 2010, \bibinfo{title}{{Viscous dissipation by tidally forced inertial modes in a rotating spherical shell},} Journal of Fluid Mechanics, 643, 363, \dodoi{10.1017/S002211200999214X}

\bibitem[{E. {Roche}(1849){Roche}}]{Roche1849}
{Roche}, E. 1849, \bibinfo{title}{{M\'emoire sur la figure d'une masse fluide, soumise l'attraction d'un point \'eloign\'e, Acad. Montpellier},} Acad. Montpellier, 1, 243

\bibitem[{N.~M. {Ros{\'a}rio} {et~al.}(2022){Ros{\'a}rio}, {Barros}, {Demangeon}, \& {Santos}}]{Rosario2022}
{Ros{\'a}rio}, N.~M., {Barros}, S.~C.~C., {Demangeon}, O.~D.~S., \& {Santos}, N.~C. 2022, \bibinfo{title}{{Measuring the orbit shrinkage rate of hot Jupiters due to tides},} \aap, 668, A114, \dodoi{10.1051/0004-6361/202244513}

\bibitem[{M. {Rozner} {et~al.}(2022){Rozner}, {Glanz}, {Perets}, \& {Grishin}}]{Rozner2022}
{Rozner}, M., {Glanz}, H., {Perets}, H.~B., \& {Grishin}, E. 2022, \bibinfo{title}{{Inflated Eccentric Migration of Evolving Gas Giants I - Accelerated Formation and Destruction of Hot and Warm Jupiters},} \apj, 931, 10, \dodoi{10.3847/1538-4357/ac6808}

\bibitem[{J. {Sanz-Forcada} {et~al.}(2011){Sanz-Forcada}, {Micela}, {Ribas}, {Pollock}, {Eiroa}, {Velasco}, {Solano}, \& {Garc{\'{\i}}a-{\'A}lvarez}}]{Sanz-Forcada2011}
{Sanz-Forcada}, J., {Micela}, G., {Ribas}, I., {et~al.} 2011, \bibinfo{title}{{Estimation of the XUV radiation onto close planets and their evaporation},} \aap, 532, A6, \dodoi{10.1051/0004-6361/201116594}

\bibitem[{K.~G. {Stassun} {et~al.}(2018){Stassun}, {Oelkers}, {Pepper}, {Paegert}, {De Lee}, {Torres}, {Latham}, {Charpinet}, {Dressing}, {Huber}, {Kane}, {L{\'e}pine}, {Mann}, {Muirhead}, {Rojas-Ayala}, {Silvotti}, {Fleming}, {Levine}, \& {Plavchan}}]{Stassun2018}
{Stassun}, K.~G., {Oelkers}, R.~J., {Pepper}, J., {et~al.} 2018, \bibinfo{title}{{The TESS Input Catalog and Candidate Target List},} \aj, 156, 102, \dodoi{10.3847/1538-3881/aad050}

\bibitem[{A.~P. {Stephan} {et~al.}(2018){Stephan}, {Naoz}, \& {Gaudi}}]{Stephan2018}
{Stephan}, A.~P., {Naoz}, S., \& {Gaudi}, B.~S. 2018, \bibinfo{title}{{A-type Stars, the Destroyers of Worlds: The Lives and Deaths of Jupiters in Evolving Stellar Binaries},} \aj, 156, 128, \dodoi{10.3847/1538-3881/aad6e5}

\bibitem[{M. {Sucerquia} \& N. {Cuello}(2025){Sucerquia} \& {Cuello}}]{Sucerquia2025}
{Sucerquia}, M., \& {Cuello}, N. 2025, \bibinfo{title}{{Extreme exomoons in WASP-49 Ab: Dynamics and detectability},} \aap, 694, L8, \dodoi{10.1051/0004-6361/202452968}

\bibitem[{D. {Tamayo} {et~al.}(2020){Tamayo}, {Rein}, {Shi}, \& {Hernandez}}]{reboundx}
{Tamayo}, D., {Rein}, H., {Shi}, P., \& {Hernandez}, D.~M. 2020, \bibinfo{title}{{REBOUNDx: a library for adding conservative and dissipative forces to otherwise symplectic N-body integrations},} \mnras, 491, 2885, \dodoi{10.1093/mnras/stz2870}

\bibitem[{T. {Tokuno} {et~al.}(2024){Tokuno}, {Fukui}, \& {Suzuki}}]{Tokuno2024}
{Tokuno}, T., {Fukui}, A., \& {Suzuki}, T.~K. 2024, \bibinfo{title}{{A Novel Method to Constrain Tidal Quality Factor from A Nonsynchronized Exoplanetary System},} \apj, 973, 128, \dodoi{10.3847/1538-4357/ad67c9}

\bibitem[{F. {Valsecchi} {et~al.}(2015){Valsecchi}, {Rappaport}, {Rasio}, {Marchant}, \& {Rogers}}]{Valsecchi2015}
{Valsecchi}, F., {Rappaport}, S., {Rasio}, F.~A., {Marchant}, P., \& {Rogers}, L.~A. 2015, \bibinfo{title}{{Tidally-driven Roche-lobe Overflow of Hot Jupiters with MESA},} \apj, 813, 101, \dodoi{10.1088/0004-637X/813/2/101}

\bibitem[{J. {Vidal} {et~al.}(2018){Vidal}, {C{\'e}bron}, {Schaeffer}, \& {Hollerbach}}]{Vidal2018}
{Vidal}, J., {C{\'e}bron}, D., {Schaeffer}, N., \& {Hollerbach}, R. 2018, \bibinfo{title}{{Magnetic fields driven by tidal mixing in radiative stars},} \mnras, 475, 4579, \dodoi{10.1093/mnras/sty080}

\bibitem[{S. {Vissapragada} {et~al.}(2022){Vissapragada}, {Chontos}, {Greklek-McKeon}, {Knutson}, {Dai}, {P{\'e}rez Gonz{\'a}lez}, {Grunblatt}, {Huber}, \& {Saunders}}]{Vissapragada2022}
{Vissapragada}, S., {Chontos}, A., {Greklek-McKeon}, M., {et~al.} 2022, \bibinfo{title}{{The Possible Tidal Demise of Kepler's First Planetary System},} ApJL, 941, L31, \dodoi{10.3847/2041-8213/aca47e}

\bibitem[{S.~M. {Wahl} {et~al.}(2021){Wahl}, {Thorngren}, {Lu}, \& {Militzer}}]{Wahl2021}
{Wahl}, S.~M., {Thorngren}, D., {Lu}, T., \& {Militzer}, B. 2021, \bibinfo{title}{{Tidal Response and Shape of Hot Jupiters},} \apj, 921, 105, \dodoi{10.3847/1538-4357/ac1a72}

\bibitem[{W. {Wang} {et~al.}(2024){Wang}, {Zhang}, {Chen}, {Wang}, {Yu}, \& {Ma}}]{Wang2024}
{Wang}, W., {Zhang}, Z., {Chen}, Z., {et~al.} 2024, \bibinfo{title}{{Long-term Variations in the Orbital Period of Hot Jupiters from Transit-timing Analysis Using TESS Survey Data},} \apjs, 270, 14, \dodoi{10.3847/1538-4365/ad0847}

\bibitem[{E.~J. {Weber} \& J. {Davis}(1967){Weber} \& {Davis}}]{Weber1967}
{Weber}, E.~J., \& {Davis}, Leverett, J. 1967, \bibinfo{title}{{The Angular Momentum of the Solar Wind},} \apj, 148, 217, \dodoi{10.1086/149138}

\bibitem[{N.~N. {Weinberg} {et~al.}(2012){Weinberg}, {Arras}, {Quataert}, \& {Burkart}}]{Weinberg2012}
{Weinberg}, N.~N., {Arras}, P., {Quataert}, E., \& {Burkart}, J. 2012, \bibinfo{title}{{Nonlinear Tides in Close Binary Systems},} \apj, 751, 136, \dodoi{10.1088/0004-637X/751/2/136}

\bibitem[{N.~N. {Weinberg} {et~al.}(2024){Weinberg}, {Davachi}, {Essick}, {Yu}, {Arras}, \& {Belland}}]{Weinberg2024}
{Weinberg}, N.~N., {Davachi}, N., {Essick}, R., {et~al.} 2024, \bibinfo{title}{{Orbital Decay of Hot Jupiters due to Weakly Nonlinear Tidal Dissipation},} \apj, 960, 50, \dodoi{10.3847/1538-4357/ad05c9}

\bibitem[{J.~N. {Winn} {et~al.}(2010){Winn}, {Fabrycky}, {Albrecht}, \& {Johnson}}]{Winn2010}
{Winn}, J.~N., {Fabrycky}, D., {Albrecht}, S., \& {Johnson}, J.~A. 2010, \bibinfo{title}{{Hot Stars with Hot Jupiters Have High Obliquities},} \apjl, 718, L145, \dodoi{10.1088/2041-8205/718/2/L145}

\bibitem[{J.~N. {Winn} {et~al.}(2018){Winn}, {Sanchis-Ojeda}, \& {Rappaport}}]{Winn2018}
{Winn}, J.~N., {Sanchis-Ojeda}, R., \& {Rappaport}, S. 2018, \bibinfo{title}{{Kepler-78 and the Ultra-Short-Period planets},} \nar, 83, 37, \dodoi{10.1016/j.newar.2019.03.006}

\bibitem[{J. {Wisdom} \& M. {Holman}(1991){Wisdom} \& {Holman}}]{wh}
{Wisdom}, J., \& {Holman}, M. 1991, \bibinfo{title}{{Symplectic maps for the N-body problem.},} \aj, 102, 1528, \dodoi{10.1086/115978}

\bibitem[{A.~S. {Wolf} \& D. {Ragozzine}(2009){Wolf} \& {Ragozzine}}]{Wolf2009}
{Wolf}, A.~S., \& {Ragozzine}, D. 2009, in IAU Symposium, Vol. 253, Transiting Planets, ed. F.~{Pont}, D.~{Sasselov}, \& M.~J. {Holman}, 163--169, \dodoi{10.1017/S1743921308026367}

\bibitem[{I. {Wong} {et~al.}(2022){Wong}, {Shporer}, {Vissapragada}, {Greklek-McKeon}, {Knutson}, {Winn}, \& {Benneke}}]{Wong2022}
{Wong}, I., {Shporer}, A., {Vissapragada}, S., {et~al.} 2022, \bibinfo{title}{{TESS Revisits WASP-12: Updated Orbital Decay Rate and Constraints on Atmospheric Variability},} \aj, 163, 175, \dodoi{10.3847/1538-3881/ac5680}

\bibitem[{I. {Wong} {et~al.}(2021){Wong}, {Shporer}, {Zhou}, {Kitzmann}, {Komacek}, {Tan}, {Tronsgaard}, {Buchhave}, {Vissapragada}, {Greklek-McKeon}, {Rodriguez}, {Ahlers}, {Quinn}, {Furlan}, {Howell}, {Bieryla}, {Heng}, {Knutson}, {Collins}, {McLeod}, {Berlind}, {Brown}, {Calkins}, {de Leon}, {Esparza-Borges}, {Esquerdo}, {Fukui}, {Gan}, {Girardin}, {Gnilka}, {Ikoma}, {Jensen}, {Kielkopf}, {Kodama}, {Kurita}, {Lester}, {Lewin}, {Marino}, {Murgas}, {Narita}, {Pall{\'e}}, {Schwarz}, {Stassun}, {Tamura}, {Watanabe}, {Benneke}, {Ricker}, {Latham}, {Vanderspek}, {Seager}, {Winn}, {Jenkins}, {Caldwell}, {Fong}, {Huang}, {Mireles}, {Schlieder}, {Shiao}, \& {Noel Villase{\~n}or}}]{Wong2021}
{Wong}, I., {Shporer}, A., {Zhou}, G., {et~al.} 2021, \bibinfo{title}{{TOI-2109: An Ultrahot Gas Giant on a 16 hr Orbit},} \aj, 162, 256, \dodoi{10.3847/1538-3881/ac26bd}

\bibitem[{J.~T. {Wright} {et~al.}(2012){Wright}, {Marcy}, {Howard}, {Johnson}, {Morton}, \& {Fischer}}]{Wright2012}
{Wright}, J.~T., {Marcy}, G.~W., {Howard}, A.~W., {et~al.} 2012, \bibinfo{title}{{The Frequency of Hot Jupiters Orbiting nearby Solar-type Stars},} \apj, 753, 160, \dodoi{10.1088/0004-637X/753/2/160}

\bibitem[{S.~W. {Yee} {et~al.}(2020){Yee}, {Winn}, {Knutson}, {Patra}, {Vissapragada}, {Zhang}, {Holman}, {Shporer}, \& {Wright}}]{Yee2020}
{Yee}, S.~W., {Winn}, J.~N., {Knutson}, H.~A., {et~al.} 2020, \bibinfo{title}{{The Orbit of WASP-12b Is Decaying},} \apjl, 888, L5, \dodoi{10.3847/2041-8213/ab5c16}

\bibitem[{L.-C. {Yeh} {et~al.}(2024){Yeh}, {Jiang}, \& {A-thano}}]{Yeh2024}
{Yeh}, L.-C., {Jiang}, I.-G., \& {A-thano}, N. 2024, \bibinfo{title}{{Searching for candidates of orbital decays among transit exoplanets},} \na, 106, 102130, \dodoi{10.1016/j.newast.2023.102130}

\bibitem[{J.~P. {Zahn}(1989){Zahn}}]{Zahn1989}
{Zahn}, J.~P. 1989, \bibinfo{title}{{Tidal evolution of close binary stars. I - Revisiting the theory of the equilibrium tide},} \aap, 220, 112

\bibitem[{J.~P. {Zahn}(2008){Zahn}}]{Zahn2008}
{Zahn}, J.~P. 2008, in EAS Publications Series, ed. M.~J. {Goupil} \& J.~P. {Zahn}, Vol.~29, 67--90, \dodoi{10.1051/eas:0829002}

\bibitem[{J. {Zendejas} {et~al.}(2010){Zendejas}, {Segura}, \& {Raga}}]{Zendejas2010}
{Zendejas}, J., {Segura}, A., \& {Raga}, A.~C. 2010, \bibinfo{title}{{Atmospheric mass loss by stellar wind from planets around main sequence M stars},} \icarus, 210, 539, \dodoi{10.1016/j.icarus.2010.07.013}

\bibitem[{V.~N. {Zharkov} \& V.~P. {Trubitsyn}(1978){Zharkov} \& {Trubitsyn}}]{Zharkov1978}
{Zharkov}, V.~N., \& {Trubitsyn}, V.~P. 1978, {Physics of planetary interiors}

\end{thebibliography}



\end{document}